\documentclass[a4paper,11pt]{article}
\usepackage[latin1]{inputenc}
\usepackage{amssymb}
\usepackage{amsmath}
\usepackage[dvips,a4paper,text={16cm,23cm},centering]{geometry}

\numberwithin{equation}{section}

\newcommand{\RR}{\mathbb{R}}

\newcommand{\NN}{\mathbb{N}}
\newcommand{\ZZ}{\mathbb{Z}}

\newcommand{\be}{\begin{equation}}
\newcommand{\ee}{\end{equation}}
\newcommand{\bea}{\begin{eqnarray}}
\newcommand{\eea}{\end{eqnarray}}
\newcommand{\ud}{\mathrm{d}}
\newcommand{\G}{\left}
\newcommand{\D}{\right}
\newcommand{\p}{\partial}

\newcommand{\cV}{\mathcal{V}}
\newcommand{\cG}{\mathcal{G}}
\newcommand{\cS}{\mathcal{S}}

\newcommand{\cQ}{\mathcal{Q}}

\newcommand{\tzeta}{\widetilde{\zeta}}
\newcommand{\fg}{\mathfrak{g}}
\newcommand{\fG}{\mathrm{G}}
\newcommand{\fh}{\mathfrak{h}}
\newcommand{\fH}{\mathrm{H}}
\newcommand{\tfg}{\widetilde{\mathfrak{g}}}
\newcommand{\tfG}{\widetilde{\mathrm{G}}}
\newcommand{\tfh}{\widetilde{\mathfrak{h}}}
\newcommand{\tfH}{\widetilde{\mathrm{H}}}
\newcommand{\ttau}{\widetilde{\tau}}

\newcommand{\w}{\wedge}

\newcommand{\cY}{\mathcal{Y}}
\newcommand{\cX}{\mathcal{X}}

\newcommand{\cW}{\mathcal{W}}
\newcommand{\cK}{\mathcal{K}}
\newcommand{\cF}{\mathcal{F}}

\newcommand{\tcV}{\widetilde{\mathcal{V}}}
\newcommand{\tcG}{\widetilde{\mathcal{G}}}
\newcommand{\tcX}{\widetilde{\mathcal{X}}}
\newcommand{\tcY}{\widetilde{\mathcal{Y}}}
\newcommand{\tXi}{\widetilde{\Xi}}
\newcommand{\tcF}{\widetilde{\mathcal{F}}}
\newcommand{\tM}{\widetilde{M}}
\newcommand{\tN}{\widetilde{N}}
\newcommand{\tLambda}{\widetilde{\Lambda}}
\newcommand{\tV}{\widetilde{V}}
\newcommand{\tv}{\widetilde{v}}
\newcommand{\tG}{\widetilde{G}}
\newcommand{\tX}{\widetilde{X}}
\newcommand{\tY}{\widetilde{Y}}
\newcommand{\tD}{\widetilde{D}}
\newcommand{\tP}{\widetilde{P}}
\newcommand{\tQ}{\widetilde{Q}}
\newcommand{\tchi}{\widetilde{\chi}}
\newcommand{\tpsi}{\widetilde{\psi}}
\newcommand{\tPsi}{\widetilde{\Psi}}
\newcommand{\tZ}{\widetilde{Z}}
\newcommand{\txi}{\widetilde{\xi}}

\newcommand{\mK}{\mathrm{K}}
\newcommand{\mE}{\mathrm{E}}
\newcommand{\mSO}{\mathrm{SO}}
\newcommand{\fk}{\mathfrak{k}}
\newcommand{\fw}{\mathfrak{w}}
\newcommand{\fe}{\mathfrak{e}}
\newcommand{\fso}{\mathfrak{so}}
\newcommand{\hsigma}{\hat{\sigma}}
\newcommand{\hpsi}{\widehat{\psi}}
\newcommand{\hSigma}{\widehat{\Sigma}}
\newcommand{\hPsi}{\widehat{\Psi}}

\begin{document}

\begin{flushright}
ULB-TH/06-06\\
April 2006\\
\vspace*{10mm}
\end{flushright}

\begin{center}
\begin{Large}
\textbf{Infinite-Dimensional Gauge Structure}

\vspace{2mm}
{\bf of d=2 N=16 Supergravity}
\end{Large}

\vspace{15mm}
{\bf Louis Paulot}

\vspace{10mm}
Physique théorique et mathématique, Université libre de Bruxelles\\
and\\
International Solvay Institutes\\
Campus Plaine C.P.~231, B--1050 Bruxelles, Belgium
\vspace{5mm}

{\ttfamily lpaulot@ulb.ac.be}
\end{center}

\vspace{30mm}
\hrule
\medskip
\begin{center}
\begin{large}
{\bf Abstract}
\end{large}
\end{center}

\begin{quote}
Dimensional reduction of maximal supergravity to two dimensions leads to an infinite-dimensional (non-local) symmetry group $\cW \ltimes \mE_9$ which has a simpler action when the bosonic fields are dualised to an infinite tower of dual potentials. We construct a doubled-valued representation of its compact subgroup $\cK \ltimes \mK(\mE_9)$ and we show that off-shell fermions take place in this infinite-dimensional representation. The equations of motion can be written in a fully gauge-covariant way as a selfduality condition for the infinite-dimensional fields. The $\cW \ltimes \mE_9$ global symmetry is thus manifest. The linear system associated to the theory is recovered in a triangular gauge. Finally we provide supersymmetry transformations for off-shell fields.
\end{quote}

\bigskip

\hrule

\newpage
\tableofcontents
\newpage

\section{Introduction}

Upon dimensional on a $n$-torus, pure gravity exhibits a $\mSO(n)$ global symmetry. When the final dimension is 3, this group is enhanced to $\mSO(n+1)$. In two dimensions, the global symmetry group contains the affine Kac-Moody extension of $\mSO(n+1)$ \cite{geroch} which is related to the integrable structure of the reduced theory \cite{maison,bz,bm2}. 
Dimensional reduction of maximal supergravity leads to enhanced symmetry groups $\mE_n$ \cite{Cremmer:1979up,Julia:1980gr} up to $E_8$ in three dimensions. In two dimensions, one gets the affine extension $\mE_9$ \cite{Julia:1981wc,bm1} which has been extended by a real form $\cW$ of the Witt group (Virasoro without central charge) \cite{Julia:1996nu,be-ju}. The symmetry group is nonlinearly realized on a tower of dual potentials which described a $\sigma$-model with an infinite-dimensional symmetric space as target space. In the vicinity of a generic spacetime singularity, maximal supergravity behaves like an infinite-dimensional coset $\mE_{10}/\mK(\mE_{10})$ and the hyperbolic Kac-Moody algebra $\mE_{10}$ has been conjectured to be a symmetry of quantum $M$-theory \cite{Damour:2000hv,Damour:2002cu,Damour:2002et,Damour:2005zb}. It has also been conjectured that the symmetry group could be the triple extension $\mE_{11}$ \cite{West:2000ga,West:2001as,Schnakenburg:2001ya,Englert:2003zs,Englert:2003py} (the link between $\mE_{10}$ and $\mE_{11}$ is studied in \cite{Englert:2004ph}) or some Borcherds superalgebras \cite{Henry-Labordere:2002dk,Henry-Labordere:2002xh}.

In any case, fermions are supposed to be in a doubled-valued representation of the maximal compact subgroup of the global symmetry group. For finite symmetry groups appearing in dimension at least three, this is established, but the representation of the compact subgroup are not classified in the infinite-dimensional case. Fermionic representations of $\mK(\mE_{10})$ have been constructed for spin 1/2 \cite{deBuyl:2005zy} and spin 3/2 \cite{deBuyl:2005mt,Damour:2005zs} with a covariant derivative giving at first levels the correct couplings to bosonic fields. However these representations are finite-dimensional and are expected to be the first levels of some infinite-dimensional representations which would be the superpartners of the infinite set of bosonic fields. A major difficulty with $\mE_{10}$ and $\mE_{11}$ is that there is no explicit description of these groups and their maximal compact subgroups. The case of $\mE_9$ is a case where the group has also infinite dimension, but with the advantage of having a description as the central extension of some loop group. It may therefore provide a good starting point for understanding further the infinite-dimensional algebraic structure of M-theory. The bosonic degrees of freedom belong to a symmetric space $(\cW \ltimes \mE_9)/(\cK \ltimes \mK(\mE_9))$ on which the global $\mE_9$ symmetry acts nonlinearly \cite{Julia:1996nu}. The theory is expressed in a fixed (triangular) gauge and the equations of motions of bosons are obtained by a Lax pair related to the integrability of the model.  In \cite{Nicolai:1988jb} fermionic terms were added to the bosonic linear system, so that the compatibility conditions give also the equations of motion for fermions. This was further studied in \cite{Nicolai:1993pe,Nicolai:1998gi}. The $\mK(\mE_9)$ structure of the theory was studied in \cite{Nicolai:2004nv} where finite-dimensional representations were constructed for fermions.

In \cite{Paulot:2004hh} a gauge-independent formulation of the bosonic sector was obtained: a gauge-covariant selfduality condition reduces the infinite number of fields to the physical ones and the usual Lax pair is derived from this constraint when the gauge is fixed. The same was achieved for $\sigma$-models with rigid supersymmetry in \cite{Paulot:2004fm}. Here we provide such a gauge-independent formulation for the full $d=2$ $N=16$ supergravity and we construct an infinite-dimensional representation of $\cK \ltimes \mK(\mE_9)$ for off-shell fermions. The maximal compact subgroup $\cK \ltimes \mK(\mE_9)$ and the representations are expressed in a way which should allow the inclusion in the larger structures of $\mE_{10}$ and $\mE_{11}$.

The structure of this paper is the following. In section \ref{section-alg} we deal with the mathematical aspects and we construct finite and infinite-dimensional representations of $\mK(\mE_9)$. Besides the representations which are very explicitely written, tools are given which allow to construct many other representations. In section \ref{flat-sigma}, the dualisation and the gauge-independent formulation of flat space $\sigma$-model is given, as it gives some pieces of the structure needed for supergravity. Finally in section \ref{section-sugra} the $d=2$ supergravity is derived as the solution of a selfduality equation on an infinite set of fields. This gives a formulation which respect the gauge structure and make the symmetry manifest. We give finally the supersymmetry transformations for off-shell fields.

\section{$\mK(\mE_9)$ representations}
\label{section-alg}

\subsection{Algebraic definitions}
\label{definitions}

$\fG$ is a real, simple Lie group and $\fH$ its maximal compact subgroup $\mK(\fG)$, fixed
pointwise by some involution $\tau$.  $\tfG$ is a real form of the
(complex) loop group on $\fG$: elements of $\tfG$ are maps $g(t)$ from the
(positive) real line to $\fG$ with some regularity conditions. Here, we request
holomorphicity in the $t$ variable in a neighbourhood of the unit
circle. (This is a dense subset of all analytic functions.) The
reality condition is that $g(t)$ must be real for $t$ real. The group
low is given by pointwise $\fG$ multiplication. $\tfH$ is
a subgroup of $\tfG$ extending $H$. Using the involution $\tau: \fG
\rightarrow \fG$ an involution on $\tfG$ can be defined:
\be
\ttau: \ g(t) \longmapsto \tau\G(g\G(\frac{1}{t}\D)\D) \rlap{\ .}
\ee
$\tfH$ is the set of elements of $\tfG$ which are invariant under
$\ttau$. It should be noted that it is not the loop group of $\fH$. We consider $\tfH$ as the ``maximal compact subgroup'' of $\tfG$.

$\fG$ has tangent (simple) Lie algebra $\fg$. we denote by the same
letter $\tau$ the involution on $\fg$ induced by $\tau: \fG
\rightarrow \fG$. It allows to decompose $\fg$ as $\fg = \fh \oplus
\fh^\perp$ where $\fh$ and $\fh^\perp$ are respectively the subspaces
of eigenvalues $+1$ and $-1$. $\fh$ is the tangent Lie algebra of $\fH$.
The Lie algebra of $\tfg$ is made of maps $\RR_+ \rightarrow \fg$ with
the same reality an regularity conditions as the Lie group $\tfG$. The
involution $\ttau: \tfG \rightarrow \tfG$ induces an involution on
$\tfg$, still denoted by $\ttau$, which gives the decomposition $\tfg
= \tfh \oplus \tfh^\perp$ on eigenspaces of eigenvalues $+1$ and
$-1$. $\tfh$ is the Lie algebra of $\tfH$ and is fixed pointwise by
$\ttau$.

The affine extension $\fg^{(1)}$ is obtained by the addition of a central charge $c$. The commutators are given by
\be
\G[ a(t) , b(t) \D] = \omega(a,b) c
\ee
where $\omega$ is an algebra 2-cocycle \cite{kac,bm2}. It can be computed by the formula
\be
\omega(a,b) = \oint_\mathcal{C} dt \langle \p_t a(t) , b(t) \rangle
\ee
where $\langle \cdot , \cdot \rangle $ is the Killing form and $\mathcal{C}$ is a contour invariant under $t \mapsto 1/t$ and avoiding poles. Usually $\mathcal{C}$ is taken to be the unit circle. Here there will be singularities in $t=\pm 1$ so $\mathcal{C}$ has to be the mean over two contours $\mathcal{C}_1$ and $\mathcal{C}_2$ exchanged by $t \mapsto 1/t$:
\be
\omega(a,b) = \frac{1}{2}\oint_{\mathcal{C}_1} dt \langle \p_t a(t) , b(t) \rangle
+ \frac{1}{2}\oint_{\mathcal{C}_2} dt \langle \p_t a(t) , b(t) \rangle
\rlap{\ .}
\ee
Concretely, we consider the sum of the residues inside the unit disc plus one half of the residues on the unit circle.

At the group level, $\fG^{(1)}$ elements are pairs $(g(t),\lambda)$ where $g(t)$ is an element of the loop group and  $\lambda$ a real number. The group law is
\be
(g_1(t),\lambda_1) (g_2(t),\lambda_2) = (g_1(t)g_2(t),\lambda_1 \lambda_2 e^{\Omega(g_1,g_2)})
\ee
where $\Omega$ is the group 2-cocycle constructed from $\omega$. A mixed cocycle
\be
\Omega'(a,g) = \lim_{\epsilon\rightarrow 0}\, \Omega(e^{\epsilon a} , g)
\ee
can also be defined \cite{bm2}.

The involution $\ttau$ defining the compact subalgebra is extended on the central charge by $c \mapsto -c$ which preserves the Lie algebra structure. As a consequence, the central charge is noncompact and therefore the ``maximal compact'' subgroup of the affine extension is the same as the ``maximal compact'' subgroup of the loop extension of $\fG$.

In this paper, $\fG$ will be the exceptional group $\mE_8$, in the split form $\mE_{8(8)}$, with its central extension $\mE_9$.

In \cite{Julia:1996nu} it was shown that bosons of maximal $d=2$ supergravity
transform as a coset
\be
\frac{\cW \ltimes \mE_9}{\cK \ltimes \mK(\mE_9)} \rlap{\ .}
\ee 
$\cW$ is formally the group of diffeomorphisms of the line. More precisely it
is defined at the Lie algebra level as a real form of the Witt algebra, 
\emph{i.e.} the complexified Virasoro algebra without central charge. It as
real generators
\be
L_n = t^{n+1} \p_t
\ee
with $n\in \NN$. This algebra will be denoted by $\fw$.
$\cK$ is its ``maximal compact subgroup''. At the Lie algebra level, $\fk$ is
the subalgebra of fixed points of $\fw$ under the involution
\be
\tau_\fw : \ L_n \ \longrightarrow \ -L_{-n} \rlap{\ .}
\ee
This subalgebra is spanned by elements $L_n - L_{-n}$.

An element of $\cW \ltimes \mE_9$ is a triple $\G( f , \tcV , \lambda \D)$ where $f$ is a real function $t=f(s)$, $\tcV$ is an element of the loop group $\widetilde{\mE_8}$ and $\lambda$ is a real number. The semi-direct product is given by the group law
\be
\G( f_1 , \tcV_1 , \lambda_1 \D)\G( f_2 , \tcV_2 , \lambda_2 \D) = \G( f_1 \circ f_2 , (\tcV_1 \circ f_2)\tcV_2 , \lambda_1 \lambda_2 e^{\Omega\!\G( \tcV_1 \circ f_2 , \tcV_2\D)} \D)
\rlap{\ .}
\ee

The ``maximal compact subgroup'' of $\cW \ltimes \mE_9$ is $\cK \ltimes \mK(\mE_9)$; it is the set of points fixed by the involution $\tau_\ltimes$ built from $\tau_\fw$ and $\ttau$.

\subsection{Single-valued infinite-dimensional representations of $\mK(\mE_9)$}

\subsubsection{$\widetilde{\fe_8} / \fk(\fe_9)$}

%In order to describe the fermionic, doubled-valued, representations of
%$\mK(\mE_9)$ needed for the maximal supergravity description, we describe here
%the single-valued representations where fermions $\tchi$ take place in the
%rigidly supersymmetric $\frac{\mE_8}{\mSO(16)/\ZZ_2}$ $\sigma$-model.\footnote{$\mE_8$ is here
%the noncompact form $\mE_{8(8)}$ of $\mE_8$. }

%In the superspace description of this model in section
%\ref{flat-sigma},\footnote{Note that it is only viewed with $N=(1,1)$ rigid
%  supersymmetry} fermions are two-dimensional fermions with value in the coset
%$\fe_8/\fso(16)$, which is precisely a 128-dimensional
%Majorana-Weyl spinorial representation of $\mathfrak{so}(16)$. This
%naturally extends to the infinitely dualised set of fields: fermions
%are two-dimensional fermions with value in the algebra coset
%$\widetilde{\fe_8} / \fk(\fe_9)$. It should be noted that by definition this
%is a single-valued representation: it can be derived from the natural action
%of $\mK(\mE_9)$ on $\widetilde{\mE_8}$. 

A infinite-dimensional representation of $\mK(\mE_9)$ is given by the linear algebra coset $\widetilde{\fe_8} / \fk(\fe_9)$. This is for example the representation where fermions $\tchi$ 
live in the rigidly supersymmetric $\sigma$-model described in section \ref{flat-sigma}.
It should be noted that by definition this
is a single-valued representation: it can be derived from the natural action
of $\mK(\mE_9)$ on $\widetilde{\mE_8}$. 

The algebra coset $\widetilde{\fe_8} / \fk(\fe_9)$
is described by a function $\tchi(t)$ with value in $\fe_8$ up to additive 
$\fk(\fe_9)$ gauge transformations
\be
\tchi(t) \ \longrightarrow \ \tchi(t) + \txi(t) \qquad \txi \in \fk(\fe_9)
\rlap{\ ,}
\label{sv-fg}
\ee
which is \emph{a priori} independent of the $\mK(\mE_9)$ adjoint
action
\be
\tchi(t) \ \longrightarrow \ \tXi(t) \tchi(t) \tXi(t)^{-1} \qquad 
\tXi \in \mK(\mE_9)
\rlap{\ .}
\label{sv-bg}
\ee
We refer to the additive $\fk(\fe_9)$ gauge freedom as ``fermionic gauge''.
A gauge choice can be made for $\tchi(t)$ with respect to $\fk(\fe_9)$. In
particular, a ``triangular'' gauge imposes $\tchi(t)$ to be regular in $t=0$;
moreover, the $t=0$ value $\tchi\!\mid_{t=0}$ must
be in $\fso(16)^\perp$. In other words, there is a usual
128-dimensional Majorana-Weyl $\fso(16)$ spinor $\chi_0$ in degree 0 and a
sequence of $\fe_8$-valued spinors $\chi_n$ for higher degrees:
\be
\tchi = \chi_0 + \sum_{n > 0} t^n \chi_n \rlap{\ .}
\ee
Under the adjoint action (\ref{sv-bg}) of $\mK(\mE_9)$, some compensating
$\fk(\fe_9)$ gauge variation may have to be added to preserve the gauge
choice. Finally, $\tchi$ transforms under $\mK(\mE_9)$
as given in (\ref{ssm-gauge-chi}):
\be
\tchi \ \longrightarrow \ \tXi \tchi \tXi^{-1} + \txi
\ee
where $\txi \in \fk(\fe_9)$ is chosen to restore the fermionic gauge
choice, which is regularity in $t=0$ here.
For an infinitesimal transformation
\be
\tXi = 1 + t^p \delta g + t^{-p} \tau(\delta g)
\ee
with $\delta g \in \fe_8$, the first part of the transformation act on
$\tchi = \sum_{n \in \ZZ} t^n \chi_n$ as
\bea
n \geq p &:& \chi_n \ \longrightarrow \ \chi_n + \G[ \delta g , \chi_{n-p}
  \D] + \G[\tau(\delta g) , \chi_{n+p} \D] \nonumber\\
0 \leq n <p &:& \chi_n \ \longrightarrow \ \chi_n + \G[\tau(\delta g)
  , \chi_{n+p} \D] \label{ke9-gauge-1}\\ 
-p \leq n < 0 &:& \chi_n \ \longrightarrow \ \G[\tau(\delta g)
  , \chi_{n+p} \D] \rlap{\ .} \nonumber\\ 
\eea
To restore the gauge $\chi_n = 0$ for $n<0$, $\txi$ must be
\be
\txi = - \sum_{0 \leq n\leq p} \G( t^{-n} \G[\tau(\delta g) ,
  \chi_{p-n} \D]  + t^n \G[ \delta g , \tau(\chi_{p-n}) \D]\D)
\ee
where the second term follows from $\tau(\txi(t^{-1})) = \txi(t)$.
Adding this to (\ref{ke9-gauge-1}) gives the complete
$\fk(\fe_9)$ action on the infinite-dimensional spinor $\tchi$:
\bea
n > p &:& \chi_n \ \longrightarrow \ \chi_n + \G[ \delta g , \chi_{n-p}
  \D] + \G[\tau(\delta g) , \chi_{n+p} \D] \nonumber\\
n=p &:& \chi_p \longrightarrow \ \chi_p + \G[ \delta g , \chi_0 -
  \tau(\chi_0) \D] + \G[\tau(\delta g) , \chi_{2p} \D]
\label{gauge-fermions-1}\\
0 \leq n <p &:& \chi_n \ \longrightarrow \ \chi_n + \G[\tau(\delta g)
  , \chi_{n+p} \D] - \G[ \delta g , \tau(\chi_{p-n})\D] 
\nonumber
\eea
With this full transformation, the gauge choice for the
infinite-dimensional fermion is preserved: $\chi_0$ is in
$\fso(16)^\perp$ and there are $\chi_n \in \fe_8$ only for $n>0$.

%(We will see in section \ref{flat-sigma} that, on-shell, 
%the fermions of the selfdual $\sigma$-model are constrained:
%\be
%\tchi_\pm = \frac{1 \mp t}{1 \pm t} \chi_\pm
%\rlap{\ .}
%\ee
%This equation is covariant with respect to $\fk(\fe_9)$ and the action
%of $\tXi = 1+ t^p \delta g + t^{-p} \tau(\delta g)$ reduces to
%\be
%\chi_\pm \ \longrightarrow \ \chi_\pm + (\mp 1)^p \G[ \delta g + \tau(\delta
%  g) , \chi_\pm \D] \rlap{\ .}
%\ee
%It means that on-shell, the infinite-dimensional spinorial
%representation of $\mK(\mE_9)$ reduces to the finite-dimensional
%spinorial representation
%\be
%\chi_\pm \ \longrightarrow \ \tXi\!\mid_{t=\mp 1} \, \chi_\pm \
%\tXi\!\mid_{t=\mp 1}^{-1}
%\ee
%acting through $\mSO(16)$ as described in (\ref{KE9-finite}).)

\subsubsection{Extended representation}

The representation just described can be extended in both (compatible) ways.
First, by adding a central charge $c$, the representation described is extended to $\fe_9/\fk(\fe_9)$. Under the action of $\tXi \in \mK(\mE_9)$, the associated field $\hpsi$ would transform as
\be
\hpsi \ \longrightarrow \ \hpsi + \Omega'(\tXi,\tchi)
\rlap{\ .}
\ee

This infinite-dimensional additive coset can be extended further to $\frac{\fw \ltimes \fe_9}{\fk \ltimes \fk(\fe_9)}$. In addition to the fields $\tchi$ and $\hpsi$, there is a field $\tpsi_{2}$ with values in the Witt algebra $\fw$,
\be
\tpsi_2 = \sum_n \psi_{2(n)} L_n = \tpsi_2^{(t)} t \p_t
\ee
defined up to\footnote{In fact, it would be consistent to take any subgroup of $\fk$ instead of $\fk$ itself. However this would result into a representation reducible with respect to $\mK(\mE_9)$: the elements which would be added by reducing the denominator of the coset would simply be trivial with respect to $\mK(\mE_9)$.} $\fk$:
\be
\tpsi_2 \sim \tpsi_2 + \tzeta
\ee
for any $\tzeta \in \fk$.

The $\mK(\mE_9)$ is defined through the natural action of this group on the algebra $\fw \ltimes \fe_9$, which preserves the subalgebra $\fk \ltimes \fk(\fe_9)$. An element $\tXi \in \mK(\mE_9)$ acts explicitely as
\bea
\tpsi_2& \longrightarrow & \tpsi_2 + \tzeta
\\
\tchi& \longrightarrow & \tchi + \tXi \tchi \tXi^{-1} - \tpsi_2^{(t)} t\p_t\tXi \tXi^{-1} + \txi
\\
\hpsi & \longrightarrow & \hpsi + \Omega'(\tXi,\tchi)
\eea
where $\tzeta$ and $\txi$ are compensating additive transformations which can be added to preserve some gauge condition.

In the triangular gauge where all fields are regular in $t=0$, the action of $\tXi = 1 + t^p \delta g + t^{-p} \tau(\delta g)$, taking into account compensating transformations, is as follows. $\tpsi_{2(n)}$ are left invariant:
\be
\psi_{2(n)} \ \longrightarrow \ \psi_{2(n)}
\label{gauge-fermions-1c}
\ee
For the field $\tchi$, (\ref{gauge-fermions-1}) is modified:
\bea
n > p &:& \chi_n \ \longrightarrow \ \chi_n + \G[ \delta g , \chi_{n-p}
  \D] + \G[\tau(\delta g) , \chi_{n+p} \D]  - p \psi_{2(n-p)} \delta g + p \psi_{2(n+p)} \tau(\delta g) \nonumber\\
n=p &:& \chi_p \longrightarrow \ \chi_p + \G[ \delta g , \chi_0 -
  \tau(\chi_0) \D] + \G[\tau(\delta g) , \chi_{2p} \D]
\label{gauge-fermions-1b}\\
0 \leq n <p &:& \chi_n \ \longrightarrow \ \chi_n - \G[ \delta g , \tau(\chi_{p-n})\D] + \G[\tau(\delta g)
  , \chi_{n+p} \D]  + p \psi_{2(p-n)} \delta g + p \psi_{2(n+p)} \tau(\delta g)
\rlap{\ .}
\nonumber
\eea

\subsection{Finite-dimensional representations of $\mK(\mE_9)$}
\label{rep-finite}

Finite-dimensional representations of $\mK(\mE_9)$ can be obtained from the
following observation: if $\tXi$ is in the extension $\tfH$ of $\fH$ as
defined in section \ref{definitions}, then $\tXi\!\mid_{t=\pm 1}$ is
in $\fH$. Indeed, an element $\tXi$ of $\tfG$ is in $\tfH$ if
\be
\tXi(t) = \tau\G( \tXi\G( \frac{1}{t} \D)\D) \rlap{\ .}
\ee
For $t=\pm 1$, this reads
\be
\tXi(\pm 1) = \tau\G( \tXi\G( \pm 1\D)\D) \rlap{\ :}
\ee
$\tXi(\pm 1)$ is a fix point of the involution $\tau$, \emph{i.e.} it
is an element of $\fH$. Therefore, any representation $R$ of $\fH$
on a vector space $F$ gives
two representations of $\tfH$ on $F$, through the composition of
representations
\be
\begin{array}{ccccc}
\tfH &\longrightarrow& \fH &\longrightarrow&  \mathrm{End}(F) \\
\tXi  &\longmapsto& \tXi\!\mid_{t=\pm 1} &\longmapsto& R\!\G(
\tXi\!\mid_{t=\pm 1}\D) \rlap{\ .}
\end{array}
\label{KE9-finite}
\ee
The maximal compact subgroup of $\fG = \mE_8$ is $\fH=\mSO(16)/\ZZ_2$.
According to what we have explained, any represention of
$\mSO(16)/\ZZ_2$ gives rise to two representations of
$\mK(\mE_9)$. Spinorial (doubled-valued) representations of $\mSO(16)/\ZZ_2$
are thus finite-dimensional spinorial representations of $\mK(\mE_9)$.

An important representation, in which the supersymmetry parameters will be taken, is given by the finite-dimensional vector representation of $\mSO(16)/\ZZ_2$ at $t=\mp 1$. We denote them by $\mathbf{16_\pm}$. Because of
the precise form of the $\ZZ_2$ quotient, these representation are doubled-valued.

More generally, finite dimensional representations can be constructed using the quasi-triangular structure of $\mK(\mE_9)$ (or more generally $\mK(\fG^{(1)})$). As $\fk(\fe_9)$ elements have to be regular in $t=\pm 1$, they can be expanded in positive powers of $u^\pm=\frac{1 \pm t}{1\mp t}$ \cite{Nicolai:2004nv}:
\be
\delta\tXi(u^\pm) = u^\pm \delta\tXi_1 + u^{\pm 2} \delta\tXi_2 + u^{\pm 3} \delta\tXi_3 + \ldots
\ee
Truncating this expansion to some given degree preserves the algebra structure. A representation of this truncation is therefore a representation of $\fk(\fe_9)$. The finite-dimensional representation we have just described corresponds to the case where only the first term is kept.

Even more generally, $u$ can be mapped to some element $U$ by a ring morphism. The truncation to some power of $u$ is the special case where $U$ satisfies $U^n = 0$.

\subsection{Infinite-dimensional fermionic representation of $\mK(\mE_9)$}
\label{rep-ferm}

Using the tensor product, other representations can be built from the building blocks described above. In particular, we consider the infinite-dimensional representation $\mathbf{16_\pm} \times
\frac{\fw \ltimes \fe_9}{\fk \ltimes \fk(\fe_9)}$. As $\mathbf{16_\pm}$ and
$\frac{\fw \ltimes \fe_9}{\fk \ltimes \fk(\fe_9)}$ are respectively
double-valued and single-valued with respect to $\mK(\mE_9)$, the product
representation is a doubled-valued representation. Explicitely, the elements of
this representation are described by 16 $\tpsi_{2\pm}^I$ in the Lie algebra defined up to the addition of 16 $\tzeta_\pm^I \in \fk$,16
$\tchi_\pm^I \in \widetilde{\fe_8}$ modulo additive $\fk(\fe_9)$ 
transformations
parametrised by 16 compact generators $\txi_\pm^I \in \fk(\fe_9)$, and finally 16 $\hpsi_\pm^I$ 
corresponding to the central charge.

Under the action of $\tXi \in K(E_9)$, $\tpsi_{2\pm}^I$, $\tchi_\pm^I$ and $\hpsi_\pm^I$ transform as
\bea
\tpsi_{2\pm}^I & \longrightarrow &
(\tXi\!\mid_{t=\mp 1})^{IJ} \ \tpsi_{2\pm}^J + \tzeta_\pm^I
\\
\tchi_\pm^I & \longrightarrow & 
(\tXi\!\mid_{t=\mp 1})^{IJ} \G( \tXi \tchi_\pm^J \tXi^{-1} - \tpsi_{2\pm}^{J(t)} t\p_t \tXi \tXi^{-1} \D) + \txi^I
\label{transfo-KE9}\\
\hpsi_\pm^I & \longrightarrow &
(\tXi\!\mid_{t=\mp 1})^{IJ} \G(\hpsi_\pm^J +\Omega'\!\G(\tXi , \tchi_\pm^J\D)\D)
\eea
where $\tzeta_\pm^I$ and $\txi_\pm^I$ are compensators to maintain the ``fermionic gauge'' of
additive $\fk \ltimes \fk(\fe_9)$ transformations. Although $\tchi_\pm^J$ is defined up to elements of
$\mathbf{16}_\pm \times \fk(\fe_9)$, $\Omega'(\tXi,\tchi_\pm^J)$ is
well-defined: $\Omega$ and therefore $\Omega'$ vanishes when both arguments
are ``compact''.

It will be convenient to use a gauge
where $\tpsi_{2\pm}^I$ and $\tchi_\pm^I$ are regular in $t=0$:
\be
\tpsi_{2\pm}^I = \sum_{n\geq 0} t^n  \psi_{2\pm(n)}^{I} t\p_t
\ee
\be
\tchi_\pm^I = \sum_{n\geq 0} t^n  \chi_{\pm(n)}^I
\ee
Each $\chi_{\pm,n}^I$ has a part invariant under $\tau$,
$\frac{1}{2} \chi_{\pm,n}^{I,JK} X^{JK}$ and an anti-invariant part
$\chi_{\pm,n}^{I,A}\, Y^A$. In addition to the regularity condition, 
we use $\txi_{\pm,0}^I \in \mathfrak{so}(16)$ to set
$\chi_{\pm , 0}^I \in \mathfrak{so}(16)^\perp$, \emph{i.e.} 
$\chi_{\pm,0}^I = \chi_{\pm,0}^{I,A}\, Y^A$.  $\tchi_\pm^I$ are thus 
described as
\be
\tchi_\pm^I = \chi_{\pm,0}^{I,A} \, Y^A + \sum_{n > 0} t^n  
\G( \frac{1}{2} \chi_{\pm , n}^{I,JK} X^{JK} + \chi_{\pm , n}^{I,A} \, Y^A \D)
\ee
where $I,J,K$ are vector indices and $A$ a spinor index with respect to
$\mathfrak{so}(16)$.

In this gauge, infinitesimal
transformations are given by (\ref{gauge-fermions-1c}) and (\ref{gauge-fermions-1b}) plus the term 
acting on the $I$ index. For $\tXi = 1+ (t^p+t^{-p}) \delta g$ with
$\delta g = \frac{1}{2} \delta g^{IJ} X^{IJ} \in \mathfrak{so}(16)$ it is
\be
\psi_{2\pm(n)}^{I} \ \longrightarrow \ \psi_{2\pm(n)}^{I} \mp 2 \delta g^{IJ} \psi_{2\pm(n)}^{J}
\ee
and
\bea
n > p &:& \chi_{\pm , n}^I \ \longrightarrow \ \chi_{\pm , n}^I 
+ \G[ \delta g \ , \ \chi_{\pm , n-p}^I 
  + \chi_{\pm , n+p}^I \D]
\mp 2 \delta g^{IJ} \chi_{\pm , n}^J
\nonumber\\
&& \qquad \qquad \qquad \qquad \qquad \qquad \qquad \qquad \qquad \qquad + p \G(-\psi_{2\pm(n-p)}^{I} + \psi_{2\pm(n+p)}^{I}\D) \delta g
\nonumber\\
n=p &:& \chi_{\pm , p}^I \longrightarrow \ \chi_{\pm , p}^I 
+ \G[ \delta g \ , \ 2\chi_{\pm , 0}^I + \chi_{\pm , 2p}^I \D]
\mp 2 \delta g^{IJ} \chi_{\pm , p}^J
\label{gauge-fermions-2}\\
0 \leq n <p &:& \chi_{\pm , n}^I \ \longrightarrow \ \chi_{\pm , n}^I 
+ \G[\delta g \ , \ \chi_{\pm , n+p}^I - \tau(\chi_{\pm , p-n}^I)\D]
\mp 2 \delta g^{IJ} \chi_{\pm , n}^J
\nonumber\\
&& \qquad \qquad \qquad \qquad \qquad \qquad \qquad \qquad \qquad \qquad 
+ p \G(\psi_{2\pm(p-n)}^{I} + \psi_{2\pm(n+p)}^{I}\D) \delta g
\nonumber
\eea
whereas for $\tXi = 1+ (t^p-t^{-p}) \delta g$ with
$\delta g = \delta g^{A} Y^A \in \mathfrak{so}(16)^\perp$ it is
\be
\psi_{2\pm(n)}^{I(t)} \ \longrightarrow \ \psi_{2\pm(n)}^{I(t)}
\ee
and
\bea
n > p &:& \chi_{\pm , n}^I \ \longrightarrow \ \chi_{\pm , n}^I
+ \G[ \delta g \ , \ \chi_{\pm , n-p}^I
  - \chi_{\pm , n+p}^I \D]
+ p \G(-\psi_{2\pm(n-p)}^{I} - \psi_{2\pm(n+p)}^{I}\D) \delta g
\nonumber\\
n=p &:& \chi_{\pm , p}^I \longrightarrow \ \chi_{\pm , p}^I
+ \G[ \delta g \ , \ 2\chi_{\pm , 0}^I -  \chi_{\pm , 2p}^I \D]
\label{gauge-fermions-3}\\
0 \leq n <p &:& \chi_{\pm , n}^I \ \longrightarrow \ \chi_{\pm , n}^I
- \G[\delta g \ , \ \chi_{\pm , n+p}^I + \tau(\chi_{\pm , p-n}^I)\D]
+ p \G(\psi_{2\pm(p-n)}^{I} - \psi_{2\pm(n+p)}^{I}\D) \delta g
\rlap{\ .}
\nonumber
\eea

The product representation described is not necessarily irreducible.
In particular, the quasi-triangular structure of $\mK(\mE_9)$ allows to perform consistent truncations at lower levels in the $u$ expansion, because lower levels cannot be reached from higher levels (which may destroy the truncation).

A general element of the representation we are interested here is not necessarily regular in $t=\pm 1$. Nevertheless, it can be written as a limit of regular elements. So let us consider a $\tchi_\pm^I$ which is regular at the fixed points $t=\pm 1$ and let us denote by $\chi_\pm^I$ its value at $t=\mp 1$: $ \chi_\pm^I = \tchi_\pm^I \! \mid_{t=\mp 1}$.
If we forget $\tpsi_{2\pm}^I$, it follows from (\ref{transfo-KE9}) that $\chi_\pm^I$ sees only 
$\tXi\!\mid_{t=\mp 1}$ which belongs to $\mSO(16)/\ZZ_2$. As a consequence, 
$\chi_\pm^I = \chi_\pm^{I,A} Y^A$ (There could be terms of the form $\chi_\pm^{I,JK} X^{JK}$ but they would be pure gauge: the linear gauge freedom $\txi\!\mid_{t=\mp 1} \in
\mathfrak{so}(16)$ can been used to set $\tchi_\pm^I\!\mid_{t=\mp 1} =
\chi_\pm^{I,A} \, Y^A \in \mathfrak{so}(16)^\perp$.) can be
restricted to an irreducible representation of $\mSO(16)/\ZZ_2$. Namely, for matter fermions of $d=2$ supergravity, we will take the $t=\mp 1$ values to be Majorana-Weyl spinor:
\be 
\chi_\pm^I = \Gamma^I_{A\dot{A}} \chi_\pm^{\dot{A}} \, Y^A
\rlap{\ .}
\ee
(In this case, the representation may be truncated further in degree 1: the mixed symmetry part can be set to zero.) If $\tpsi_{2\pm}^I$ is nonvanishing, it generates a term in the variation of $\tchi_\pm^I$ which does not preserve the truncation. However, because of the triangular structure of $\mK(\mE_9)$, it can be safely projected to preserve the truncation. We denote by
\be
\G( \tpsi_{2\pm}^{I(t)} t\p_t \tXi \tXi^{-1}\D)_T
\ee
this projection.

If there are poles at the fixed points $t=\pm 1$, the function can be seen as the limit of regular functions. As a consequence, the truncation can be performed on the higher order pole at the fixed point $t=\mp 1$. (However, this means that we do not consider addition of elements in the representation, but only the action of the group $\mK(\mE_9)$. If one want to have an additive structure, the truncation can only be imposed on some fixed maximal degree pole.)

%As
%$\tXi \in \mK(\mE_9)$ and $\txi \in \fk(\fe_9)$ are regular functions of 
%$t$ on the unit circle, the transformation (\ref{transfo-KE9}) preserves 
%the poles of $\tchi_\pm^I(t)$ on this circle. More precisely, the higher order
%pole in each point is preserved by $\mK(\mE_9)$. The subspace of
%$\mathbf{16_\pm} \times \frac{\widetilde{\fe_8}}{\fk(\fe_9)}$ 
%made of meromorphic
%functions $\tchi_\pm^I(t)$ with fixed higher order poles on the unit circle is
%thus preserved by $\mK(\mE_9)$.

%(However, if one wants a vector space
%representation, with an additive law, only a set of maximal poles can be
%defined: addition can reduce the order of poles and even suppress
%them. As addition of fermions does not seem to be useful, we may consider an
%action of $\mK(\mE_9)$ on a space without additive structure, as for bosons
%which are parametrised by an element of the coset space
%$\frac{\mE_8}{\mSO(16)/\ZZ_2}$, which is not a vector space.)

%If one wants the representation vector space to be complete, one cannot
%impose conditions on meromorphic singularities, as a singular function can be
%reached as the limit of a sequence of regular functions in $t$. Note that this
%is not specific to the fermionic representations.

%If the singularities are fixed, the representation can be
%reduced further on the higher degree poles as described above, with $\chi_\pm^I (t \pm 1)^{-n}$ the more %singular term of $\tchi_\pm^I$ in $t=\mp 1$.

\subsection{Action of $\cK$}
\label{k-action}

There is a natural action of $\fw$ and therefore $\fk$ on all $t$-dependent fields.
When acting on $\fe_9$, $\fk$ preserves the decomposition $\fe_9 = \fk(\fe_9) \oplus \fk(\fe_9)^\perp$; gauge elements in $\fk(\fe_9)$ remain in $\fk(\fe_9)$ and there is thus a well-defined induced action on the coset $\fe_9/\fk(\fe_9)$. It can be remarked that ``compact'' generators $L_n - L_{-n} = -(t^n - t^{-n}) t\p_t$ vanish at $t=\pm 1$, so that the $t=\pm 1$ values of an element of $\mK(\mE_9)$ are left unchanged under the action of $\cK$. As a consequence, the truncation performed in section \ref{rep-ferm} is preserved. Moreover, the finite-dimensional representation defined through this special points do not see directly the action of $\cK$. Therefore, fermions in $\mathbf{16_\pm} \times (\fe_9/\fk(\fe_9)) $ or one of its subrepresentations are in the same representation of $\fk$, given by the action $L_n - L_{-n}$ on $t$. 

The scalar field $\rho$ is promoted to an element of the coset $\cW/\cK$ \cite{Julia:1996nu}. The field strength has values in the Lie algebra coset $\fw/\fk$: it is an element of $\fw$
\be
\ud f \!\circ\! f^{-1} \p_t = \sum A_n L_n
\ee
defined up to $\fk$ gauge transformations. The supersymmetric partners $\tpsi_{2\pm}^I$ are fermions in the representations $\mathbf{16_\pm} \times (\fw/\fk)$:
\be
\tpsi_{2\pm}^I = \sum_{n \in \ZZ} \tpsi_{2\pm,n}^{I(t)} t^n \p_t 
\ee
up to additive $\fk$ transformations
\be
\tpsi_{2\pm}^I \ \longrightarrow \ \tpsi_{2\pm}^I + k_n (t^n - t^{-n}) t\p_t \rlap{\ .}
\ee
As for other fields, the triangular gauge is defined as regularity in $t=0$, \emph{i.e.} vanishing of terms of negative degree in the $t$ expansion.

For all fields, the bare action of $\cK$ can break the gauge choice; in this
case, it must be followed by compensting gauge transformations which restores
the gauge.

The action of $\cK$ and $\mK(\mE_9)$ are combined into a semi-direct product $\cK \ltimes \mK(\mE_9)$, because $\cK$ acts on $\mK(\mE_9)$. We have remarked that the $t=\pm 1$ values of an element of $\mK(\mE_9)$ are left unchanged under the action of $\cK$, so that the finite-dimensional representation defined through this special values do not see directly the action of $\cK$. However, if some $\mK(\mE_9)$  compensating transformation is needed to fix the gauge on some field like $\tcV$, the action of $\cK$ will be seen even on finite-dimensional representations through this compensator.

\section{Flat space $\sigma$-model}
\label{flat-sigma}

The $d=2$ flat space\footnote{We will use here Lorentzian signature
  but results can be adapted to Euclidean space.}
$\sigma$-model on a symmetric space
$\fG/\fH$ can be seen as a $\sigma$-model on an
infinite-dimensional symmetric space $\tfG/\tfH$ constrained by a
selfduality condition. \cite{Paulot:2004fm}

\subsection{Pure bosonic model}

\subsubsection{Description of the model}

The bosonic $\sigma$-model is given by a field from
2-dimensional spacetime to the target space $\fG/\fH$. Practically,
one considers a field $\cV(x)$ in $\fG$, with gauge group
$\fH$. The field strength
\be
\cG = \ud \cV \cV^{-1}
\ee 
is a $\fg$-valued 1-form and can be decomposed as
\be
\cG = Q + P
\ee
with $Q \in \fh$ and $P \in \fh^\perp$.

From its definition, $\cG$ satisfies a Bianchi identity, which is the
pull back of the Maurer-Cartan equation:
\be
\ud \cG = \cG \w \cG \rlap{\ .}
\ee
Decomposed on $\fh \oplus \fh^\perp$, it reads
\bea
\ud Q &=& Q \w Q + P \w P \\
\ud P &=& \G[ Q , P \D]
\rlap{\ .}
\eea
This latter equation can be rewritten as
\be
\nabla P = 0
\ee
with the $\fH$-covariant derivative
\be
\nabla = \ud - \G[Q , \, \cdot \, \D]
\rlap{\ .}
\ee

Under a gauge transformation
\be
\cV(x) \longrightarrow \Xi(x)\cV(x)
\ee
with $\Xi(x) \in \fH$, $P$ is covariant whereas $Q$ transforms as a
gauge connection:
\bea
P & \longrightarrow & \Xi P \Xi^{-1} \\
Q & \longrightarrow & \ud \Xi \Xi^{-1} + \Xi Q \Xi^{-1}
\rlap{\ .}
\eea
The field strength $\cG$ is invariant under a global, right action of
$\fG$:
\be
\cV(x) \longrightarrow \cV(x) \Lambda
\ee
with $\Lambda \in \fG$. If the gauge is fixed, such a transformation
can break the gauge fixing condition which must be restored by a
local, left action of the gauge group $\fH$. In a fixed
gauge, global symmetries act thus on the field strength through
compensating gauge transformations.

The dynamics of the $\sigma$-model is governed by the action
\be
S = \frac{1}{2} \int \langle P , *\!P \rangle 
\ee
where $\langle \cdot , \cdot \rangle$ is a $\fH$-invariant symmetric form.
This gives the equation of motion
\be
\nabla\!*\!P = 0 \rlap{\ .}
\ee

\subsubsection{Selfduality constraint}
\label{bsm-deriv}

Let us consider a field in $\tfG / \tfH$, with a representative 
$\tcV(x) \in \tfG$; two configurations are
equivalent if they can be related by a gauge transformation
\be
\tcV \longrightarrow \tXi \tcV
\ee
with $\tXi \in \tfH$.
More precisely, the field lives in the adherence of $\tfG / \tfH$:
poles on the unit circle are allowed. A crucial point is that
the gauge group is $\tfH$, without singularities on the unit circle.
Otherwise, the formalism which follows would describe a null theory.

The loop algebra field strength $\tcG$ can be decomposed on $\tfh
\oplus \tfh^\perp$ as
\be
\tcG = \cX + \cY
\ee
with $\cX$ and $\cY$ respectively invariant and anti-invariant under
$\ttau$.
Expanding $\tcG$ as
\be
\tcG = \sum_{n \in \ZZ} t^n A_n
\label{boso-flat-g}
\ee
with $A_n \in \fg$, we have
\bea
\cX &=& \frac{1}{2} \sum_{n \in \ZZ} \G( t^n A_n + t^{-n} \tau(A_n) \D)
\\
\cY &=& \frac{1}{2} \sum_{n \in \ZZ} \G( t^n A_n - t^{-n} \tau(A_n) \D)
\rlap{\ .}
\label{boso-flat-y}
\eea

We define a selfduality operator $\cS$ on $\tfg$ as
\be
\cS: \ t^n \alpha \longrightarrow - t^{1-n} \tau(*\alpha) 
\label{boso-flat-sd}
\ee
where $\alpha$ is a $\fg$-valued 1-form. This allows to consider the
selfduality constraint
\be
\cS \cY = \cY \rlap{\ .}
\ee
It can be shown \cite{Paulot:2004fm} that
\begin{enumerate}
\item this constraint is covariant with respect to $\tfh$ gauge
  transformations;
\item the solutions of this selfduality constraint are the classical
  solutions of the $\fG/\fH$ $\sigma$-model. 
\end{enumerate}
With this description, the infinite dimensional symmetry $\tfG$ is
manifest: it is simply the symmetry group of the infinite dimensional
symmetric space $\tfG/\tfH$.

To see that the selfduality constraint gives the $\fG/\fH$
$\sigma$-model, one must fix partially the gauge to a triangular
configuration: $\tcV$ has to be regular in the vicinity of $t=0$.
In this gauge, $\tcG=\ud \tcV\tcV^{-1}$ is also regular in $t=0$: 
$A_n=0$ for $n<0$ in (\ref{boso-flat-g}). Due to the regularity in
$t=0$, $\cV = \tcV(t=0)$ is well defined and lives in $\fG/\fH$. Its
derivative gives $\cG = \ud \cV \cV^{-1} = A_0$ which is decomposed on 
$\fh \oplus \fh^\perp$ as $\cG = Q+P$. Solutions to (\ref{boso-flat-sd2})
are given by
\be
A_n = 2 *^n P
\ee
for $n>0$. (In Lorentzian 2-dimensional space, the Hodge star squares
to the identity: $*^2 = 1$.) It gives the following expansion for $\tcG$:
\be
\tcG = Q +P + 2t *\!P + 2t^2 P + 2t^3 *\!P + 2t^4 P + \ldots
\ee
After resummation, this gives for the
$\tfG/\tfH$ field strength
\be
\tcG = Q + \frac{1+t^2}{1-t^2} P + \frac{2t}{1-t^2} *\!P \rlap{\ ,}
\label{boso-flat-lax}
\ee
which is the linear system (Lax pair) associated to the flat space
$\sigma$-model. In a formal way, this can be rewritten as
\be
\tcG = Q + \frac{1+t*}{1-t*}P \rlap{\ .}
\ee
In lightcone coordinates $x^\pm$, the Hodge star acts on 1-forms as
\be
*: \
A_\pm \longrightarrow \mp A_\pm
\ee
and the linear system (\ref{boso-flat-lax}) reads
\be
\p_\pm \tcV \tcV^{-1} = Q_\pm + \frac{1\mp t}{1 \pm t} P_\pm
\rlap{\ .}
\ee

The Bianchi identity for $\tcG$, $\ud \tcG = \tcG \w
\tcG$ gives the Bianchi identities and the equation of motion for the 
$\fG/\fH$ $\sigma$-model:
\bea
\ud Q - Q \w Q &=& P \w P \\
\nabla P &=& 0 \\
\nabla\!*\!P &=& 0
\eea
where $\nabla$ is still the $\fH$ covariant derivative $\nabla = \ud -
\G[ Q,\,\cdot\,\D]$.

\subsubsection{Gauge structure}

For this flat space model, the gauge covariance is easy to
check. For $\cY$ satisfies $\ttau(\cY) = - \cY$, \emph{i.e.} $\cY(t)
= -\tau\G(\cY\G(\frac{1}{t}\D)\D)$, the selfduality constraint
(\ref{boso-flat-sd}) can be rewritten as
\be
\cY = t *\!\cY \rlap{\ .}
\label{boso-flat-sd2}
\ee
As left action of $\tfH$ commutes with shift of $t$-degree, this
equation is indeed covariant with respect to such gauge transformations.

Let us analyse in more details the $\tfH$ gauge structure of the
theory. In a general gauge, the equation
(\ref{boso-flat-sd2}) can be rewritten in components as
\be
A_n - \tau(A_{-n}) = *^n \G( A_0 - \tau(A_0)\D) \rlap{\ .}
\label{boso-flat-sd-compo}
\ee
Under an infinitesimal gauge transformation with parameter 
$\delta \tXi_p = t^p \delta \xi_p + t^{-p} \tau(\delta \xi_p) \in \tfh$, $\tcV$
transforms as
\be
\tcV \longrightarrow \tcV + \delta  \tXi_p \tcV
\ee
so that $\tcG = \ud \tcV \tcV^{-1}$ transforms as
\be
\tcG \longrightarrow \tcG + \ud \delta  \tXi_p + \G[ \delta \tXi_p , \tcG
  \D] \rlap{\ .}
\ee
For the $t$ expansion, this reads
\be
A_n \longrightarrow A_n + \delta_{n,p} \ud \delta \xi_p + \delta_{n,-p}
\ud \tau(\delta \xi_p) + \G[ \delta \xi_p , A_{n-p}\D] + \G[
  \tau(\delta \xi_p) , A_{n+p}\D]
\rlap{\ .}
\label{boso-flat-An}
\ee
This transformation preserves (\ref{boso-flat-sd-compo}), with
\be
A_0 - \tau(A_0) \longrightarrow A_0 - \tau(A_0) + 
\G[\delta \xi_p + \tau(\delta \xi_p), *^p (A_0 + \tau(A_0)) \D]
\rlap{\ .}
\label{boso-flat-A0}
\ee

When $\tfG$ acts on $\tcV$ on the right, it can break the gauge
choice. In triangular gauge, with only nonnegative powers of $t$ in
$\tcV$, elements of $\tfG$ with negative powers of $t$ break the
regularity in $t=0$ and one has to act on the left by a gauge
transformation to recover the triangular gauge. Such gauge
transformations belong to a very limited class of $\tfH$: they must
preserve the condition $A_n=0$ for $n<0$. These are what are called
``on-shell transformations'' in \cite{Nicolai:2004nv}:
they preserve the form
of the linear system (\ref{boso-flat-lax}). More precisely, they
correspond to $\tfH$ gauge transformations which can be seen as
compensating transformations for the action of $\tfG$ in a fixed,
triangular gauge.
Consider indeed an infinitesimal $\tfh$ gauge transformation with
parameter
\be
\delta \tXi = \sum_{p\in \NN} \G( t^p \delta \xi_p + t^{-p} \tau(\delta
\xi_p) \D)
\ee
which preserves the property $A_n =0$ for $n<0$.  By summation of
(\ref{boso-flat-An}) for negative $n$, this means that
\be
\ud \delta \tXi + \G[ \delta \tXi , \tcG \D]
\ee
must not have poles in $t=0$: only nonnegative powers of $t$ are allowed:
\be
\tcV \ud\!\G( \tcV^{-1} \delta \tXi \tcV \D) \tcV^{-1} =
\ud \delta \tXi + \G[ \delta \tXi , \tcG \D] = \delta \phi(x,t)
\ee
for some function $\delta \phi(x,t)$ holomorphic in $t=0$.
As $\tcV$ is also holomorphic in $t=0$, this is also the case for
$\tcV^{-1} \delta\phi \tcV = \ud\!\G( \tcV^{-1} \delta \tXi \tcV \D)$.
It follows that $\tcV^{-1} \delta \tXi \tcV$ can be written as
\be
\tcV^{-1} \delta \tXi \tcV = \delta\tLambda(t) + \delta \Phi(x,t)
\ee
where $\tLambda(t)$ is constant but may have negative degree
components in its $t$-expansion whereas $\Phi$ has no negative
degree (with respect to $t$) component but is generically not 
constant in $x$. The gauge transformation
\be
\tcV \longrightarrow (1 + \delta \tXi) \tcV
\ee
can thus be written as
\be
\tcV \longrightarrow \tcV + \tcV \delta \tLambda + \tcV \delta \Phi
\rlap{\ .}
\ee
It means that any  gauge transformation of this kind can be seen as
the compensator for a global transformation
\be
\tcV \longrightarrow \tcV (1 - \delta \tLambda)
\rlap{\ :}
\ee
it kills precisely the negative degree part of this transformation. It
induces also a transformation of the physical fields
\be
P \longrightarrow P + \sum_{p \in \NN} \G[ \delta \xi_p + \tau(\delta
  \xi_p) , *^p P \D]  
\rlap{\ ,}
\label{boso-flat-dP}
\ee
where the $\tau(\delta \xi_p)$ are the negative degree components in 
the expansion of $\tcV \delta \tLambda \tcV^{-1}$. 
In lightcone coordinates, (\ref{boso-flat-dP}) reads
\be
P_\pm \longrightarrow P_\pm + \G[ \delta \tXi(\mp 1),P\pm \D]  
\rlap{\ .}
\ee

\subsection{Supersymmetric model}

In view of what we intend to study in following sections ($N=16$
supergravity), we consider now supersymmetric models with at least $N=(1,1)$
supersymmetry. (Results are not hard to adapt for chiral 
supersymmetry, see \cite{Paulot:2004fm}.)

In superspace with fermionic coordinates $\theta^+$ and $\theta^-$,
the supersymmetry generators are
\be
\cQ_\pm = \frac{\p}{\p{\theta^\pm}} - \theta^\pm \p_\pm
\ee
and the supercovariant derivatives are
\be
D_\pm = \frac{\p}{\p{\theta^\pm}} + \theta^\pm \p_\pm
\rlap{\ .}
\ee
These definitions gives anticommutation relations
\be
\begin{array}{rclrcl}
\G[ \cQ_\pm , \cQ_\pm \D] &=& -\p_\pm \qquad &
\G[ \cQ_\pm , \cQ_\mp \D] &=& 0 \\
\G[ D_\pm , D_\pm \D] &=& \p_\pm &
\G[ D_\pm , D_\mp \D] &=& 0 \\
\G[ \cQ_\pm , D_\pm \D] &=& 0 &
\G[ \cQ_\pm , D_\mp \D] &=& 0 \rlap{\ .}
\end{array}
\ee

\subsubsection{Description of the model}

We consider a set of even superfields encoded in 
$V(x^\pm,\theta^\pm) \in \fG/\fH$:
\be
V = \G(1+ \theta^+ \chi_+ + \theta^- \chi_- + \theta^+ \theta^- (v +
\chi_- \chi_+)\D) \cV
\ee
where $\cV$ is a representative in $\fG$ of its class in $\fG/\fH$ and
$\psi_\pm$ and $v$ belong similarly to the Lie algebra $\fg$.
The super-field strength is given by the pair of odd superfields
\be
G_\pm = D_\pm V V^{-1}
\label{Gpm-def}
\ee
which can be decomposed on $\fh\oplus\fh^\perp$ as $G_\pm = X_\pm +
Y_\pm$. 
$X_\pm$ is the $\fh$ gauge connection and and $Y_\pm$ the
$\fH$-covariant field strength.
Under a gauge transformation
\be
V(x,\theta) \ \longrightarrow \ Z(x,\theta) V(x,\theta)
\label{ssm-gauge}
\ee
they transform indeed as
\bea
X_\pm &\longrightarrow&  D_\pm Z Z^{-1} + Z X_\pm Z^{-1} \\
Y_\pm  &\longrightarrow& Z Y_\pm Z^{-1} \rlap{\ .}
\eea

From its definition, $G_\pm$ satisfies a supersymmetric version of the 
Bianchi identity\footnote{For odd fields as $G_\pm$, brackets are of course
\emph{anticommutators}.},
\be
D_+ G_- + D_- G_+ = \G[G_+ , G_- \D] \rlap{\ ,}
\ee
which after projection on $\fh$ and $\fh^\perp$ reads
\bea
\G( D_+ Y_- - \G[ X_+ , Y_- \D] \D) + \G( D_- Y_+ - \G[ X_- , Y_+ \D] \D)
&=& 0 \label{ssm-bianchi-1}\\
D_+ X_- + D_- X_+ - \G[ X_+ , X_- \D] &=& \G[Y_+,Y_-\D] \rlap{\ .}
\label{ssm-bianchi-2}
\eea

The supersymmetric extension of the bosonic $\sigma$-model is governed
by the action
\be
\mathcal{S} = \int \ud x^2 \ud\theta^2 \ \langle Y_+ , Y_- \rangle
\rlap{\ .}
\ee
The equation of motion, written in superspace,
\be
\G(D_+ Y_- - \G[ X_+ , Y_- \D] \D) - \G( D_- Y_+ - \G[ X_- , Y_+ \D] \D) = 0
\label{ssm-eom}
\ee
is similar to the purely bosonic one written in lightcone coordinates.

In components, the field strength $G_\pm$ can be computed from its
definition (\ref{Gpm-def}):
\bea
G_+ = && \chi_+ \nonumber\\
&+& \theta^+ \G( \p_+ \cV \cV^{-1} + \chi_+ \chi_+ \D) \nonumber\\
&+&
\theta^- \G( v + \G[ \chi_+ , \chi_- \D] \D) \nonumber\\
&+& \theta^+ \theta^- \G( \p_+ \chi_- - \G[ \p_+ \cV \cV^{-1} ,
  \chi_- \D] + \G[ v + \G[ \chi_+ , \chi_- \D] , \chi_+ \D] \D)\\ 
G_- = && \chi_- \nonumber\\
&-& \theta^+ v \nonumber\\
&+& \theta^- \G( \p_- \cV \cV^{-1} + \chi_- \chi_- \D) \nonumber\\
&+& \theta^+ \theta^- \G( - \p_- \chi_+ + \G[ \p_- \cV \cV^{-1} ,
  \chi_+ \D] + \G[ v , \chi_- \D] \D) \rlap{\ .}
\eea
In addition to the decomposition
\be
\p_\pm \cV \cV^{-1} = Q_\pm + P_\pm \rlap{\ ,}
\ee
$\chi_\pm$ and $v$ must also be decomposed on $\fh \oplus \fh^\perp$ as
\bea
\chi_\pm &=& \chi_\pm^\fh + \chi_\pm^\perp \nonumber\\
v &=& v^\fh + v^\perp \rlap{\ .}
\eea 
in order to compute the connection part $X_\pm$ and the covariant part $Y_\pm$
of $G_\pm = X_\pm + Y_\pm$.
With such notations, the equation of motion (\ref{ssm-eom}) reads in
components
\bea
\G( \p_+ P_- - \G[ Q_+ + \chi_+^\perp \chi_+^\perp , P_- \D] \D) + \G( \p_-
P_+ - \G[ Q_- + \chi_-^\perp \chi_-^\perp , P_+ \D] \D) &=& 0 
\label{ssm-cV-eom}\\
\p_- \chi_+^\perp - \G[ Q_- + \chi_-^\perp \chi_-^\perp , \chi_+^\perp \D]
&=& 0
\label{ssm-chi+-eom}\\
\p_+ \chi_-^\perp - \G[ Q_+ + \chi_+^\perp \chi_+^\perp , \chi_-^\perp \D]
&=& 0
\label{ssm-chi--eom}\\
v^\perp &=& - \G[ \chi_-^\perp , \chi_+^\fh \D]
\label{ssm-v-eom}
\eea
$v^\perp$ is an auxiliary field which can be forgotten. $\chi_\pm^\fh$ and
$v^\fh$ are not involved in the equations of motion of physical degrees of
freedom: they completely decouple and are in fact pure gauge: under a gauge 
transformation (\ref{ssm-gauge}) with parameter
\be
Z(x,\theta) = \G(1 + \theta^+ \xi_+(x) + \theta^- \xi_-(x) + \theta^+
\theta^- \G(z(x) + \xi_-(x)\xi_+(x)\D)\D) \Xi(x) \rlap{\ ,}
\label{ssm-gauge-p}
\ee
fields transform as
\bea
\cV &\longrightarrow& \Xi \cV \\
\chi_\pm  &\longrightarrow& \Xi \chi_\pm \Xi^{-1} + \xi_\pm 
\label{ssm-gauge-chi}\\
v &\longrightarrow& \Xi v \Xi^{-1} - \G[ \xi_+ ,  \Xi \chi_- \Xi^{-1} \D] +
z
\label{ssm-gauge-2}
\eea
where $\Xi$ is in gauge group $\fH$ whereas $\xi_\pm$ and $v$ lie 
in the Lie algebra 
$\fh$. It follows that $\chi_\pm^\fh$ and $v^\fh$ can be set to zero by
gauge transformations which do not transform other fields:
only $\Xi$ acts on physical degrees of freedom, $\xi_\pm$ and $z$ can be
used to annihilate $\chi_\pm^\fh$ and $v^\fh$, such that
$\chi_\pm=\chi_\pm^\perp$ and $v=v^\perp$ lie in $\fh^\perp$. 
(From the equation of motion 
(\ref{ssm-v-eom}), the auxiliary field $v$ vanishes completely in this
case.) We thus recover the usual set of components variables for the
supersymmetric $\sigma$-model: in addition to the scalars $\cV$, there are
spinors $\chi_\pm$ with value in $\fh^\perp$, which is a representation of
$\fh$ through the Lie bracket in $\fg$.

The model has a global (geometric) symmetry $\fG$, which acts on $V$
by right multiplication:
\be
V \ \longrightarrow \ V \Lambda
\ee
with $\Lambda \in \fG$. In the parametrisation used here, where $V= (1
+ ... ) \cV$, only $\cV$ is transformed
\be
\cV \ \longrightarrow \ \cV \Lambda
\ee
while other fields are left invariant.
Such a transformation can break the gauge choice which must be restored
through a left gauge transformation, leading to the full nonlinear
realization of the symmetry 
\be
V(x,\theta)  \ \longrightarrow \ Z(x,\theta) V(x,\theta) \Lambda
\ee
where the gauge parameter $Z \in \fH$ depends on $V$ and $\Lambda$.
In a gauge where $\chi_\pm^\fh = 0$ and $v^\fh=0$, the gauge
parameter does not depend on $\theta$: $Z = \Xi(x)$, where $\Xi(x)$
is a gauge transformation in $\fH$ which restores the gauge choice for
$\cV$. In this case, the nonlinear symmetry acts on components as
\bea
\cV &\longrightarrow& \Xi \cV \Lambda \nonumber \\
\chi_\pm & \longrightarrow& \Xi \chi_\pm \Xi^{-1} \nonumber \\
v & \longrightarrow& \Xi v \Xi^{-1} \rlap{\ .}
\eea

In addition, there is global supersymmetry, with generators $Q_\pm$
defined above. With odd parameters $\epsilon^\pm$, it acts on
components 
as
\bea
\cV &\longrightarrow& \cV + \epsilon^+ \chi_+ \cV + \epsilon^- \chi_- \cV \\
\chi_+ & \longrightarrow& \chi_+ + \epsilon^+ \G( \p_+ \cV\cV^{-1} +
\chi_+ \chi_+ \D) + \epsilon^- \G( v + \G[ \chi_- , \chi_+ \D]\D)\\
\chi_-  & \longrightarrow& \chi_- - \epsilon^+ v + \epsilon^- \G( \p_-
\cV\cV^{-1} + \chi_- \chi_- \D)\\
v & \longrightarrow& v - \epsilon^+ \p_+ \chi_- + \epsilon^- \G( \p_-
\chi_+ - \G[ \p_-\cV\cV^{-1} , \chi_+ \D] - \G[ v , \chi_- \D] \D)
\label{ssm-susy-v}
\rlap{\ .}
\eea
This transformation breaks the (partial) gauge $\chi_\pm^\fh=0$,
$v^\fh=0$. It must therefore be followed by a gauge transformation
(\ref{ssm-gauge}) which restores these conditions.
In particular, restoring $\chi_+^\fh = 0$ requires the parameter
$\xi_+$ of (\ref{ssm-gauge-p}) to be
\be
\xi_+ = -\delta \chi_+^\fh = - \epsilon^+ \G( Q_+ + \chi_+ \chi_+ \D)
- \epsilon^- \G[ \chi_- , \chi_+ \D] \rlap{\ .}
\ee
This value plays a role in the gauge transformation (\ref{ssm-gauge-2})
for $v^\perp$: there is a additional term $-\G[ \xi_+ , \chi_-\D]$
in (\ref{ssm-susy-v}) coming from this gauge
restoration. After restoration of the partial gauge $\chi_\pm^\fh=0$,
$v^\fh=0$, the full supersymmetry transformations are thus
\bea
\cV &\longrightarrow& \cV + \epsilon^+ \chi_+ \cV + \epsilon^- \chi_- \cV \\
\chi_+ & \longrightarrow& \chi_+ + \epsilon^+ P_+ + \epsilon^- v\\
\chi_-  & \longrightarrow& \chi_- - \epsilon^+ v + \epsilon^- P_-\\
v & \longrightarrow& v - \epsilon^+ \G( \p_+ \chi_- \G[ Q_+ + \chi_+
  \chi_+ , \chi_- \D] \D) + \epsilon^- \G( \p_-
\chi_+ - \G[ Q_- + \chi_- \chi_-, \chi_+ \D] \D)
\rlap{\ .}
\eea
If a gauge has been fixed for $\cV \in \fG/\fH$, it is also
necessary to add a compensating gauge transformation $\Xi\in\fH$ acting on
$\cV$.

\subsubsection{Supersymmetric linear system}

A linear system can associated to the supersymmetric model on the same
lines as for the bosonic case. One must consider a superfield
\be
\tV = \G( 1 + \theta^+ \tchi_+ + \theta^- \tchi_- + \theta^+\theta^-
\G( \tv + \tchi_-\tchi_+ \D) \D) \tcV
\ee
with value in the infinite-dimension symmetric space
$\tfG/\tfH$. All fields depends on the spectral parameter $t$, with
$\tcV(t)$ in the loop group $\tfG$ and $\tchi_\pm(t)$ and $\tv(t)$ in
the loop algebra $\tfg$. Two field configurations are equivalent if
they are related by a gauge transformation $\tZ \in \tfH$:
\be
\tV \ \longrightarrow \ \tZ\tV \rlap{\ .}
\ee
The components transformations are given in previous section, with
$\fH$ replaced by $\tfH$: one has just to add twiddles to formulas.

The super-field strength is derived from $\tV$: $\tG_\pm = D_\pm \tV
\tV^{-1}$ and is decomposed on $\tfh \oplus \tfh^\perp$ as $G_\pm =
\tX_\pm + \tY_\pm$; $\tX\pm$ is the $\tfh$ gauge connection while
$\tY$ is the covariant field strength. The $\fh^\perp$ part is

The selfduality operator $\cS$ is modified in the following way:
\be
\cS: \ t^n A_\pm(t) \longrightarrow \mp t^{1-n} \tau\G(A_\pm(t)\D)
\rlap{\ .}
\ee
As in the bosonic case, the selfduality equation
\be
\cS \tY_\pm = \tY_\pm
\label{ssm-sd}
\ee
reduces the infinite-dimensional set of fields $\tV$ to the
finite-dimensional $\fG/\fH$ $\sigma$-model. The derivation is the
same as in the bosonic case (see section \ref{bsm-deriv}). 
First, we choose a triangular gauge for $\tV(t)$: it must be
regular in $t=0$. The $t=0$ value of this superfield,
$V=\tV\!\mid_{t=0}$ belongs to the coset $\fG/\fH$. 
As for the bosonic case, the selfduality constraint
gives the following form for the Taylor expansion of $\tG_\pm$:
\be
\tG_\pm = X_\pm + Y_\pm \mp 2t Y_\pm + 2t^2 Y_\pm \mp 2t^3 Y_\pm + \ldots
\ee
which gives the meromorphic expression for the field strength
\be
\tG_\pm = X_\pm + \frac{1\mp t}{1\pm t} Y_\pm \rlap{\ .}
\label{ssm-sdsol}
\ee
$X_\pm$ and $Y_\pm$ are respectively the $\fh$ and $\fh^\perp$ parts of
$G_\pm = D_\pm V V^{-1}$. $X_\pm$ and $Y_\pm$ are thus the $t=0$ values of
respectively $\tX$ and $\tY$. 
The Bianchi identity
\be
D_+ \tG_- + D_- \tG_+ = \G[ \tG_+ , \tG_- \D]
\ee
gives the Bianchi identities
(\ref{ssm-bianchi-1})--(\ref{ssm-bianchi-2}) 
for the $\fG/\fH$ $\sigma$-model and the equation of motion (\ref{ssm-eom}).

$\tG_\pm$ reads in components
\bea
\tG_+ = && \tchi_+ \nonumber\\
&+& \theta^+ \G( \p_+ \tcV \tcV^{-1} + \tchi_+ \tchi_+ \D) \nonumber\\
&+&
\theta^- \G( \tv + \G[ \tchi_+ , \tchi_- \D] \D) \nonumber\\
&+& \theta^+ \theta^- \G( \p_+ \tchi_- - \G[ \p_+ \tcV \tcV^{-1} ,
  \tchi_- \D] + \G[ \tv + \G[ \tchi_+ , \tchi_- \D] , \tchi_+ \D] \D)\\ 
G_- = && \tchi_- \nonumber\\
&-& \theta^+ \tv \nonumber\\
&+& \theta^- \G( \p_- \tcV \tcV^{-1} + \tchi_- \tchi_- \D) \nonumber\\
&+& \theta^+ \theta^- \G( - \p_- \tchi_+ + \G[ \p_- \tcV \tcV^{-1} ,
  \tchi_+ \D] + \G[ \tv , \tchi_- \D] \D) \rlap{\ .}
\eea

The first component of equation (\ref{ssm-sdsol}) is
\be
\tchi_\pm = \chi_\pm^\fh + \frac{1\mp t}{1\pm t} \chi_\pm^\perp
\rlap{\ .}
\label{ssm-tchi}
\ee
The degree zero gauge freedom can be used to set $\chi_\pm^\fh=0$. It
must be noted that higher degree components of $\tchi_\pm$ have
off-shell both $\fh$ and $\fh^\perp$ components: this is
(\ref{ssm-tchi}) which set all higher degree $\fh$ components to zero 
on-shell, in triangular gauge. (In non-triangular gauge, there can be
such higher degree $\fh$ components in $\tchi_\pm$.)
In this very particular gauge,
(\ref{ssm-tchi}) becomes
\be
\tchi_\pm = \frac{1\mp t}{1\pm t} \chi_\pm
\rlap{\ .}
\ee

(\ref{ssm-sdsol}) contains also the linear system for $\tcV$:
\be
\p_\pm \tcV \tcV^{-1} + \tchi_\pm \tchi_\pm = \G( Q_\pm + \chi_\pm^\fh
\chi_\pm^\fh +  \chi_\pm^\perp \chi_\pm^\perp \D) + \frac{1 \mp t}{1 \pm
  t} \G( P_\pm + \G[ \chi_\pm^\fh , \chi_\pm^\perp \D] \D)
\label{ssm-tcV}
\ee
which gives, after inclusion of (\ref{ssm-tchi}),
\be
\p_\pm \tcV \tcV^{-1} = Q_\pm + \frac{1 \mp t}{1 \pm
  t} P_\pm \pm \frac{4t}{(1\pm t)^2} \chi_\pm^\perp \chi_\pm^\perp
\rlap{\ ,}
\label{ssm-tcV-ho}
\ee
where the gauge field $\chi_\pm^\fh$ does not appear.
An important fact is the appearance of a term of higher order in
$u=\frac{1-t}{1+t}$ or $u^{-1}$, although (\ref{ssm-tcV}) is
only of order 1 in $u^{\pm 1}$. This is the inclusion of an other order 1 
equation which is responsible for higher order terms. In the gauge
$\chi=\chi^\perp$, (\ref{ssm-tcV}) is simpler:
\be
\p_\pm \tcV \tcV^{-1} + \tchi_\pm \tchi_\pm = \G( Q_\pm +
\chi_\pm \chi_\pm \D) + \frac{1 \mp t}{1 \pm
  t} P_\pm
\rlap{\ .}
\ee
As in the bosonic case, the Bianchi identity for $\tcV$ leads to the
equation of motion for $\cV$ (\ref{ssm-cV-eom}).

There are two equations for $\tv$ in (\ref{ssm-sdsol}), one coming
from $\tG_+$ and the other from $\tG_-$:
\bea
\tv + \G[ \tchi_+ , \tchi_- \D] &=& \G( v^\fh + \G[ \chi_+^\fh ,
  \chi_-^\fh \D] + \G[ \chi_+^\perp , \chi_-^\perp \D] \D) +
\frac{1-t}{1+t}\G( v^\perp +  \G[ \chi_+^\fh , \chi_-^\perp \D] + \G[
  \chi_+^\perp , \chi_-^\fh \D] \D) \label{ssm-ls-v1}\\
\tv &=& v^\fh + \frac{1+t}{1-t} v^\perp \label{ssm-ls-v2}\rlap{\ .}
\eea
The substraction between these two lines, after replacing
$\tchi_\pm$ by (\ref{ssm-tchi}), gives
\be
\G(\frac{1-t}{1+t} - \frac{1+t}{1-t} \D) \G( v^\perp + \G[
  \chi_-^\perp , \chi_+^\fh \D]\D) =0 \rlap{\ ,}
\label{ssm-tv-eom}
\ee
which leads to the equation (\ref{ssm-v-eom}) 
for the auxiliary field $v^\perp$. Finally, the solution is thus
\be
\tv = \tv^\fh - \frac{1+t}{1-t}\G[ \chi_-^\perp , \chi_+^\fh \D] 
\label{ssm-tv}
\ee
where $\tv^\fh$ and $\chi_+^\fh$ are pure gauge.
In the gauge $\chi_\pm^\fh=0$ and
$v^\fh=0$, the linear system (\ref{ssm-ls-v1})-(\ref{ssm-ls-v2})
is just
\bea
\tv + \G[ \tchi_+ , \tchi_- \D] &=& \G[ \chi_+ ,
  \chi_- \D] +
\frac{1-t}{1+t} \ v  \\
\tv &=& \frac{1+t}{1-t} \ v
\eea
and leads to $v=0$, and therefore $\tv=0$.

For $\tchi_+$, there are two equations in the linear system,
(\ref{ssm-tchi}) and
\begin{align}
 \p_- \tchi_+ - \G[ \p_- \tcV \tcV^{-1} , \tchi_+ \D] - \G[ \tv ,
   \tchi_- \D] = &\G( \p_- \chi_+^\fh - \G[ Q_- , \chi_+^\fh \D] - \G[ P_-
   , \chi_+^\perp \D] - \G[ v^\fh , \chi_-^\fh \D] - \G[ v^\perp ,
   \chi_-^\perp\D] \D) \nonumber\\
+ \frac{1+t}{1-t} &\G( \p_- \chi_+^\perp - \G[
   Q_- , \chi_+^\perp \D] - \G[ P_- , \chi_+^\fh \D] - \G[ v^\fh ,
   \chi_-^\perp \D] - \G[v^\perp , \chi_-^\fh \D] \D) \rlap{\ .}
\end{align}
Using expressions (\ref{ssm-tchi}), (\ref{ssm-tcV-ho}) and
(\ref{ssm-tv}) for $\tchi_\pm$, $\p_- \tcV\tcV^{-1}$ and $v$ in this
equation leads to
\be
\G( \frac{1-t}{1+t} - \frac{1+t}{1-t}\D) \G( \p_- \chi_+^\perp - \G[
  Q_- + \chi_-^\perp \chi_-^\perp , \chi_+^\perp \D] \D) = 0 \rlap{\ .}
\ee
This gives precisely the equation of motion (\ref{ssm-chi+-eom}) for
$\chi_+$.
Similarly, combining the linear system 
\be
\p_+ \tchi_- - \ldots = (\ldots) + \frac{1-t}{1+t} (\ldots)
\ee
with
\be
\tchi_- = \chi_-^\fh + \frac{1+t}{1-t} \chi_-^\perp
\ee
leads to the equation of motion
(\ref{ssm-chi--eom}) for $\chi_-$.
As for other fields, the equations simplify in the gauge
$\chi_\pm^\fh=0$, $v^\fh=0$:
\begin{align}
\p_- \tchi_+ - \G[ \p_- \tcV \tcV^{-1} , \tchi_+ \D] - \G[ \tv ,
   \tchi_- \D] &= \G( - \G[ P_- , \chi_+ \D] - \G[ v ,
   \chi_-\D] \D) + \frac{1+t}{1-t} \G( \p_- \chi_+ - \G[
   Q_- , \chi_+ \D]\D) \\
\p_+ \tchi_- - \G[ \p_+ \tcV \tcV^{-1} , \tchi_- \D] - \G[ \tv + \G[
   \tchi_+, \tchi_- \D] ,
   \tchi_+ \D] &= \G( - \G[ P_+ , \chi_- \D] - \G[ v + \G[ \chi_+ ,
   \chi_- \D], \chi_+\D] \D) \nonumber\\ &\qquad\qquad\qquad
+ \frac{1-t}{1+t} \G( \p_+ \chi_- - \G[
   Q_+ , \chi_- \D]\D) \ . 
\end{align}

In this gauge, we have seen that the solution to the linear system for
the auxiliary field is $\tv = 0$. Taking this into account, the linear
systems for other fields can be rewritten as
\bea
\p_+ \tcV \tcV^{-1} + \tchi_+ \tchi_+ = \G( Q_+ +
\chi_+ \chi_+ \D) + \frac{1 - t}{1 + t} P_+
\label{rs-ls-1}
\\
\p_- \tcV \tcV^{-1} + \tchi_- \tchi_- = \G( Q_- +
\chi_- \chi_- \D) + \frac{1 + t}{1 - t} P_-
\eea
\bea
\tchi_+ &=& \frac{1-t}{1+t} \, \chi_+ \\
\p_- \tchi_+ - \G[ \p_- \tcV \tcV^{-1} , \tchi_+ \D] 
   &=& - \G[ P_- , \chi_+ \D] + \frac{1+t}{1-t} \G( \p_- \chi_+ - \G[
   Q_- , \chi_+ \D]\D) 
\eea
\bea
\p_+ \tchi_- - \G[ \p_+ \tcV \tcV^{-1} , \tchi_- \D] 
   &=& - \G[ P_+ , \chi_- \D] + \frac{1-t}{1+t} \G( \p_+ \chi_- - \G[
   Q_+ , \chi_- \D]\D) \\
\tchi_- &=& \frac{1+t}{1-t}\, \chi_-
\rlap{\ .}
\label{rs-ls-6}
\eea
As we have already remarked, there are only terms of order 1 in
$u^{\pm 1} = \G( \frac{1-t}{1+t}\D)^{\pm 1}$, 
but replacing $\tchi_\pm$ by their explicit expression in $t$
produces higher order terms for $\p_\pm \tcV \tcV^{-1}$.

\section{Linear system for supergravity}
\label{section-sugra}

\subsection{$d=2$ supergravity}

Two-dimensional $N=16$ supergravity \cite{Nicolai:1987vy} can be obtained by
dimensional reduction from higher dimension maximal supergravity, and
especially from $D=11$ supergravity \cite{Cremmer:1978km}. In three
dimensions, maximal $N=16$ supergravity \cite{Marcus:1983hb} has a
$E_8$-symmetric structure. The gravitational field, the dreibein, is not 
dynamical in three dimensions. Other bosonic degrees of freedom
are scalars which define a $\sigma$-model on the noncompact symmetric space
$\frac{E_{8(8)}}{Spin(16)/\ZZ_2}$. Supersymmetric partners are gravitini in
the vector representation of $SO(16)$ and
matter fermions in a Majorana-Weyl representation of $SO(16)$, with opposite
chirality with respect to matter bosons.

In two dimensions, the dreibein split into a zweibein, a dilaton $\rho$ and
a Kaluza-Klein 1-form  $B$. 
Moreover, coordinate choice and Lorentz invariance are used
to bring the zveibein in diagonal form (conformal gauge). Finally, the
dreibein reads
\be
(e_\mu{}^a) = \G( 
\begin{array}{cc}
\lambda \delta_\mu^a & \rho B_\mu \\
0 & \rho
\end{array} \D)
\ee
The Kaluza-Klein 1-form is auxiliary in two dimensions and
bring quartic spinor terms upon elimination \cite{Nicolai:1988jb}; we
therefore do not consider it in this paper.
Similarly, the gravitini $\psi_a^I$ is split into two-dimensional
gravitini $\psi_a^I$ and the third component $\psi_2^I$. Supersymmetry is
used to bring the gravitini in superconformal gauge $\psi_a^I = \gamma_a
\psi^I$, analogous to the diagonal form of the zweibein. $\psi^I$ and
$\psi_2^I$ are thus superpartners of respectively $\lambda$ and $\rho$.

Matter bosons are scalars which parametrise a symmetric space,
\be
\cV \in \frac{E_{8(8)}}{Spin(16)/\ZZ_2} \rlap{\ .}
\label{coset-e8}
\ee 
The field strength
$\cG = \ud \cV \cV^{-1}$
is a 1-form with values in the algebra $\mathfrak{e}_8$ and can be
decomposed along 
$\mathfrak{e}_8 = \mathfrak{so}(16) \oplus \mathfrak{so}(16)^\perp$
as $\cG = Q + P$. $Q$ is the gauge vector and $P$ the covariant field
strength. The $\mathfrak{so}(16)$ subalgebra is spanned by elements 
$X^{IJ}$ with usual commutation relations. 
Its orthogonal complement $\mathfrak{so}(16)^\perp$, with
generators $Y^A$, has dimension
128 and behaves as a Majorana-Weyl representation with respect
to $SO(16)$:
\be
\G[ X^{IJ} , Y^A \D] = -\frac{1}{2} \Gamma^{IJ}_{AB} Y^B \rlap{\ .}
\label{XijYa}
\ee
The algebra is closed with commutation relations for noncompact generators
\be
\G[ Y^A , Y^B \D] = \frac{1}{4} \Gamma^{IJ}_{AB} X^{IJ} \rlap{\ .}
\ee
(A proof of the Jacobi identity for these commutators can be found in
\cite{deWit:1992up}.)
In this basis, the field strength reads
\be
\cG = Q + P = \frac{1}{2} Q^{IJ} X^{IJ} + P^A Y^A
\ee
with indices $I,J$ running from 1 to 16 and $A$ from 1 to 128. From
(\ref{XijYa}) the action of $\mathfrak{so}(16)$ on the covariant field
strength components reads
\be
X^{IJ} \cdot P^A = \frac{1}{2} \Gamma^{IJ}_{AB} P^B \rlap{\ .}
\ee

Matter fermions have opposite chirality with respect to $\mathfrak{so}(16)$
and are parametrised as $\chi^{\dot{A}}$. They transforms under the
$\mathfrak{so}(16)$ action as
\be
X^{IJ} \cdot \chi^{\dot{A}} = \frac{1}{2} \Gamma^{IJ}_{\dot{A}\dot{B}}
\chi^{\dot{B}} \rlap{\ .}
\ee
The $\ZZ_2$ quotient in (\ref{coset-e8}) is such that bosons are
single-valued with respect to the gauge action whereas fermions are
double-valued.

After a rescaling of fermions by a factor $(-i\lambda)^{\frac{1}{2}}$, the
equations of motion can be put in the following form, up to higher order
fermionic terms \cite{Nicolai:1998gi}.
\bea
\begin{array}{r}
\rho^{-1} D_-\!\G( \rho \G( P_+^A - 2 \psi_{2+}^I \Gamma^I_{A\dot{A}}
\chi_+^{\dot{A}} \D) \D) \\ +  
\rho^{-1} D_+\!\G( \rho \G( P_-^A + 2 \psi_{2-}^I \Gamma^I_{A\dot{A}}
\chi_-^{\dot{A}} \D) \D)
\end{array}
&=& \begin{array}{l}
\G( -\frac{1}{4} \chi_-^{\dot{A}} 
\Gamma^{IJ}_{\dot{A}\dot{B}} \chi_-^{\dot{B}} +2 \psi_{2-}^I \psi_-^J 
\D) \Gamma^{IJ}_{AB} P_+^B \\
+ \G( -\frac{1}{4} \chi_+^{\dot{A}}
\Gamma^{IJ}_{\dot{A}\dot{B}} \chi_+^{\dot{B}} -2 \psi_{2+}^I \psi_+^J
\D)  \Gamma^{IJ}_{AB} P_-^B
\end{array}
\label{eom1}
\\
\rho^{-\frac{1}{2}} D_\pm\!\G( \rho^{\frac{1}{2}}\chi_\mp^{\dot{A}} \D)
&=& \mp \frac{1}{2} \psi_{2\mp}^I \Gamma_{A\dot{A}}^I P_\pm^A
\label{eom2}
\\ 
\p_+ \p_- \rho &=& 0
\label{eom3}
\\
D_\pm \!\G( \rho \psi_{2\mp}^I\D) &=& 0 
\label{eom4}
\\
\p_+\p_- \sigma &=& -\frac{1}{2} P_+^A P_-^A +  \G( \chi_+^{\dot{A}} D_-
\chi_+^{\dot{A}} + \chi_-^{\dot{A}} D_+\chi_-^{\dot{A}} \D)
\label{eom5}
\\
D_\pm \psi_\mp^I &=& -\frac{1}{2} P_\pm^A \Gamma^I_{A\dot{A}}
\chi_\mp^{\dot{A}}
\label{eom6}
\\
\rho^{-1} \p_\pm \rho \, \p_\pm \hat\sigma &=& \frac{1}{2} P_\pm^A P_\pm^A 
+ \chi_\pm^{\dot{A}} D_\pm \chi_\pm^{\dot{A}} 
\pm \psi_{2\pm}^I \Gamma^I_{A\dot{A}} \chi_\pm^{\dot{A}} P_\pm^A \nonumber\\
&& \pm \psi_\pm^I \rho^{-1} D_\pm\!\G( \rho \psi_{2\pm}^I \D) 
\pm \psi_{2\pm}^I D_\pm\psi_\pm^I
\label{eom7}
\\
\mp \rho^{-1} D_\pm\!\G( \rho \psi_{2\pm}^I \D) &=& \mp \p_\pm \sigma \,
\psi_{2\pm}^I + \rho^{-1} \p_\pm \rho
\, \psi_\pm^I - P_\pm^A \Gamma^I_{A\dot{A}} \chi_\pm^{\dot{A}}
\label{eom8}
\eea
where $\sigma = \ln(\lambda)$ and $\hat\sigma = \sigma -
\frac{1}{2}\ln(\p_+\rho \p_-\rho)$, which behaves like a genuine scalar.
$D_\pm$ are covariant derivatives with respect to $\mathfrak{so}(16)$:
\bea
D_\pm P^A &=& \p_\pm P^A + \frac{1}{4} Q_\pm^{IJ} \Gamma^{IJ}_{AB} P^B \\
D_\pm \chi^{\dot{A}} &=& \p_\pm \chi^{\dot{A}} + \frac{1}{4} Q_\pm^{IJ} 
\Gamma^{IJ}_{\dot{A}\dot{B}} \chi^{\dot{B}} \\
D_\pm \psi^I &=& \p_\pm \psi^I + Q_\pm^{IJ} \psi^J \rlap{\ .}
\eea
(\ref{eom1}) and (\ref{eom2}) are the equations of motion of matter. 
(\ref{eom3}) to (\ref{eom6}) are the equations for the gravitational sector.
(\ref{eom7}) and (\ref{eom8}) are respectively the conformal and
superconformal constraints; they are obtained through the variation of
gravitational fields which are set to zero in superconformal gauge, the
diagonal, traceless part of the metric and the gravitino. The first order
equation (\ref{eom7}) together with other equations of motion implies 
the second order equation for the conformal factor (\ref{eom5}).

Reparametrisation and local supersymmetry are used to fix the superconformal
gauge: there remains only superconformal invariance. Conformal
transformations are parametrised by chiral functions $f^+(x^+)$ and
$f^-(x^-)$ ($\p_\mp f^\pm = 0$).
Scalars have weight 0, spacetime derivatives weight 1,
whereas fermions have weight $\frac{1}{2}$:
\bea
\rho(x^+,x^-) &\longrightarrow& \rho(f^+(x^+),f^-(x^-)) \\
\p_\pm \rho &\longrightarrow&  (\p_\pm f^\pm) \, \p_\pm\rho \\
\psi_\pm  &\longrightarrow&  (\p_\pm f^\pm)^{\frac{1}{2}} \, \psi_\pm 
\eea
and similarly for other field (recall $\hat{\sigma}$ behaves as a scalar but
not $\sigma$).
In the same way, equations are invariant under local supersymmetry
transformations with chiral parameters $\epsilon^I_\pm$ such that $D_\mp
\epsilon_\pm^I = 0 $ (up to cubic fermionic terms):
\be
\begin{array}{rclcrcl}
\delta_\pm \cV \cV^{-1} &=& - 2 \epsilon_\pm^I \Gamma^I_{A\dot{A}}
\chi_\pm^{\dot{A}} Y^A 
&\quad &
\delta_\pm \chi_\pm^{\dot{A}} &=& \epsilon_\pm^I \Gamma^I_{A\dot{A}}
P_\pm^A
\vspace{2mm} \\
\delta_\pm \rho  \rho^{-1} &=& \pm 2 \epsilon_\pm^I \psi_{2\pm}^I
&&
\delta_\pm \psi_{2\pm}^I  &=& \mp \epsilon_\pm^I \p_\pm \rho \rho^{-1}
\vspace{2mm} \\
\delta_\pm \sigma &=& - 2 \epsilon_\pm^I \psi_\pm^I
&&
\delta_\pm \psi_\pm^I  &=& D_\pm \epsilon_\pm^I + \p_\pm \hsigma
\epsilon_\pm^I \rlap{\ .}
\end{array}
\label{susy-finite}
\ee
The usual supersymmetry transformations are in fact
\be
\delta^{usual}_\pm(\epsilon_\pm) = \sqrt{\p_\pm \rho} \ \delta^{here}_\pm\!\G(\frac{\epsilon_\pm} {\sqrt{\p_\pm \rho}} \D)
\ee
which replaces $\p_\pm \hsigma$ is replaced by $\p_\pm \sigma$ in the last line of (\ref{susy-finite}).

\subsection{Fields}

In analogy with the rigid supersymmetric case linear system
(\ref{rs-ls-1})--(\ref{rs-ls-6}) and from what is already known in the
supersymmetric case, we consider fields with an infinite number of
components. The matter scalars
$\cV(x) \in \frac{E_8}{Spin(16)/\ZZ_2}$ are extended to an element of the
infinite-dimensional symmetric space:
$\tcV \in \widetilde{\mE_8}/\mK(\mE_9)$. ($\widetilde{E_8}$ is the loop
group extension of $E_8$. $K(\widetilde{E_8})=K(E_9)$ is its maximal
``compact'' subgroup. See section \ref{definitions}.)

From supersymmetry transformations (\ref{susy-finite}), the gravitini
$\psi_\pm^I$ can be considered as a gauge field for supersymmetry
variations, and they are superpartners for the conformal factor. As this
field is not promoted to a $t$-dependent field in the purely bosonic case
\cite{Julia:1996nu,Paulot:2004hh}, gravitini $\psi_\pm^I$ and therefore
supersymmetry parameters $\epsilon_\pm^I$ are also kept $t$-independent.
This means that the supersymmetry parameters $\epsilon_\pm^I$ are taken in
the finite-dimensional, doubled-valued representations $\mathbf{16_\pm}$ described in section \ref{rep-finite}. This gives two distinct representations of $\mK(\mE_9)$
for left and right-handed $\epsilon_\pm^I$ (and $\psi_\pm^I$).

From the supersymmetry transformations, if matter bosons are in
$\frac{\mE_8}{\mSO(16)/\ZZ_2}$ and supersymmetry generators in
$\mathbf{16_\pm}$, then matter bosons must lie in $\mathbf{16_\pm} \times \frac{\mE_8}{\mSO(16)/\ZZ_2}$. 
Matter
fermions are thus infinitely ``dualised'' to an infinite tower of fields $\tchi_\pm^I$.
More precisely $\tchi_\pm^I$ are taken in the truncated infinite-dimensional representations of $K(E_9)$ which are described in section \ref{rep-ferm}, where there is a Majorana-Weyl fermion at the first level in $u^\pm=\frac{1 \pm t}{1 \mp t}$.

The $\widetilde{\mE_8}/\mK(\mE_9)$ coset describing matter bosons is in fact
extended to a $\mE_9/\mK(\mE_9)$ symmetric space when the conformal factor is
considered \cite{Julia:1981wc,bm2}. 
In the absence of fermions, the central charge
is coupled to a field $\hsigma = \sigma - \frac{1}{2} \ln({\p_+\rho}) -
\frac{1}{2} \ln({\p_-\rho})$. Including fermions, we define the field 
associated to the central charge $\hsigma$ to be, up to higher order fermionic
terms,
\be
\hsigma = \sigma - \frac{1}{2} \ln({\p_+\rho}) - \frac{1}{2} \ln({\p_-\rho})
- \frac{\rho \psi^I_+ \psi^I_{2+}}{\p_+\rho}  - \frac{\rho \psi^I_-
  \psi^I_{2-}}{\p_-\rho} \rlap{\ .}
\ee
It has been seen in section \ref{rep-ferm} that under the action of $\tXi \in \mK(\mE_9)$ 
on the coset $\mE_9/\mK(\mE_9)$, this field
transforms as
\be
\hsigma \ \longrightarrow \ \hsigma + \Omega(\tXi,\tcV)
\ee
where $\Omega$ is the group 2-cocycle defining the central extension
\cite{bm2}. 

%The supersymmetric variation gives
%\be
%\delta_\pm \hsigma = -2 \epsilon_\pm^I \G( \hpsi^I_\pm + \Omega'\G( \tcV^{-1} , \tchi_\pm^I \D) \D)
%\ee
%with
Its supersymmetric partner is parametrised as
\be
\hpsi^I_\pm = \frac{1}{2} \psi^I_\pm + \frac{1}{2} \frac{\p_\pm \sigma \,
  \rho \psi^I_{2\pm}}{\p_\pm \rho} - \frac{1}{2} \frac{D_\pm(\rho
  \psi^I_{2\pm})}{\p_\pm \rho} - \frac{1}{2} \frac{\langle P_\pm , \tchi_\pm^I \rangle}{\rho^{-1} \p_\pm \rho}
  \rlap{\ ,}
\ee
up to higher order fermionic terms.
This fermionic field is combined with matter fermions $\tchi^I_\pm$ into a
subspace of $\mathbf{16}_\pm \times ( \fe_9/\fk(\fe_9))$ (the conditions
which can be imposed on $\tchi^I_\pm$ to reduce the representation are 
not modified). Thus $\hpsi^I_\pm$ transforms under $\tXi \in \mK(\mE_9)$ as
\be
\hpsi^I_\pm \ \longrightarrow \ (\tXi\!\mid_{t=\mp 1})^{IJ} \G(\hpsi^J_\pm 
+ \Omega'(\tXi,\tchi_\pm^J) \D)
\ee
where $\Omega'$ is the mixed cocycle introduced in
\cite{bm2}. 

The dilaton $\rho$ is extended to an element of the coset $\cW/\cK$, \emph{i. e.} to a
function $t=f(s)$ modulo the left action of diffeomorphisms $k(t)$ preserving the unit circle \cite{Julia:1996nu}:
\be
f \ \longrightarrow \ k \circ f \rlap{\ .}
\ee
There is a natural right action of $\cW$ defined by composition on the right:
\be
f \ \longrightarrow \ f \circ g \rlap{\ .}
\ee
The field f depends on spacetime position, and $\cK$ is a gauge group with local action.
The derivative $\ud f \circ f^{-1}$ is a 1-form with
values in the Witt algebra coset $\fw/\fk$. It is invariant under the global right action of $\cW$. As for matter bosons, a gauge can be fixed and this would generate compensating $\cK$ transformations on $\ud f \circ f^{-1}$.

Its supersymmetric partners are
$t$-dependent fermions $\tpsi_{2\pm}^I$ with values in the Witt algebra. We
can decompose them as 
\be
\tpsi_{2\pm}^I = \tpsi_{2\pm}^{I(t)} t\p_t
\ee
or
\be
\tpsi_{2\pm}^I = \tpsi_{2\pm}^{I(u)} u^{\pm} \p_{u^\pm}
\ee
with $u^\pm = \frac{1\pm t}{1\mp t}$. 
$\tpsi_{2\pm}^{I(t)}$ and $\tpsi_{2\pm}^{I(u)}$
are related by
\be
\tpsi_{2\pm}^{I(u)} = \frac{\pm 2 t}{1 - t^2} \tpsi_{2\pm}^{I(t)}
\rlap{\ .}
\ee 
Naively, the additive gauge would be the ``maximal compact subalgebra'' $\fk$, acting as
\be
\tpsi_{2\pm}^I \ \longrightarrow \ \tpsi_{2\pm}^I + \tzeta_{\pm}^I
\ee
for $\tzeta_{\pm}^I \in fk$. In fact, the equations of motion will only be invariant under the subgroup $\fk_\pm \subset \fk$ of functions $\tzeta_{\pm}^{I(t)}(t)$ with vanishing derivative, in the variable $t$, at $t=\mp 1$. As a consequence, the usual triangular gauge condition with only positive powers of $t$ cannot be imposed; one can however impose a gauge condition with an expansion in $t$ starting at degree $-1$:
\be
\tpsi_{2\pm}^{I(t)} = \frac{1}{t} \psi_{2\pm,-1}^{I(t)} + \psi_{2\pm,0}^{I(t)} + t \psi_{2\pm,1}^{I(t)} + t^2 \psi_{2\pm,2}^{I(t)} + \ldots
\ee
(In this case, $\tpsi_{2\pm}^I = \tpsi_{2\pm}^{I(t)} t\p_t$ is regular in $t$: it does not introduce singularities in $t=0$ when acting on a regular function of $t$.)

It was shown in \cite{Julia:1996nu} that all bosons of maximal $d=2$ supergravity
form a coset with a semi-direct structure coming from the action of reparametrisation of the $t$ variable on $t$-dependent matter fields:
\be
\G( f , \tcV , e^{\hsigma} \D) \in \frac{\cW \ltimes \mE_9}{\cK \ltimes \mK(\mE_9)} \rlap{\ .}
\ee 
In the same way, fermions are in the truncated representation
\be
\G( \tpsi_{2\pm} , \tchi_\pm , \hpsi_\pm \D) \in \mathbf{16}_\pm \times_T \G(\frac{\fw \ltimes \fe_9}{\fk_\pm \ltimes \fk(\fe_9)} \D) 
\ee 
where $T$ denotes the truncation of the direct product described in section \ref{definitions}. Because of the presence of $\fk_\pm$ instead of $\fk$ in the denominator, this is in fact a reducible representation of $\mK(\mE_9)$ and $\mK \ltimes \mK(\mE_9)$: it is the direct sum of the irreducible representation described in section \ref{rep-ferm} and one copy of the trivial representation
\be
\mathbf{16}_\pm \times_T \G(\frac{\fw \ltimes \fe_9}{\fk_\pm \ltimes \fk(\fe_9)} \D) = 
\mathbf{16}_\pm \times_T \G(\frac{\fw \ltimes \fe_9}{\fk \ltimes \fk(\fe_9)} \D) \oplus \mathbf{1}
\rlap{\ .}
\ee
The additional field will in fact be related to supersymmetry variations, studied in section \ref{section-susy}.

\subsection{Selfduality constraint}

In \cite{Paulot:2004hh}, it was shown that the linear system of $d=2$ supergravity can be generalised in a  $\cK \ltimes \mK(\mE_9)$-gauge covariant way as a selfduality equation. We give here the fermionic corrections to bosonic equations and the parallel selfduality constraints for fermionic fields.

A field strength $(\tcF_\pm , \tcG_\pm , \hSigma_\pm)$ is defined as
\be
\tcF_\pm = \p_\pm f \!\circ\! f^{-1} \p_t 
+ \G[ \tpsi_{2\pm}^I , -2\tpsi_{2\pm}^I \D] 
\ee
\be
%\begin{split}
\tcG_\pm = \p_\pm \tcV \tcV^{-1}
%&
+ \G[ \p_\pm f \!\circ\! f^{-1} \p_t, \tcV \D] \tcV^{-1}
+ \G[ \tpsi_{2\pm}^I , -2 \tchi_\pm^I \D]
%\\
%&
+ \frac{1}{4} \langle \tchi_\pm^I , \tchi_\pm^J \rangle X^{IJ}
%- 2\tpsi_{2\pm}^{I(u)} \psi_\pm^J X^{IJ}
-4 \tpsi_{2\pm}^{I(u)} \tpsi_{2\pm}^{J(u)} X^{IJ}
%\end{split}
\ee
\be
\hSigma_\pm = \p_\pm \hsigma - \Omega'\!\G(\tcV^{-1},\tcG_\pm \D) - \omega\!\G( \tchi_\pm^I , \tchi_\pm^I \D)
\rlap{\ .}
\ee
The bilinear form $\langle \cdot , \cdot \rangle$ in the definition of $\tcG_\pm$ is the $\mE_8$ Killing form, acting pointwise with respect to $t$: the term $\frac{1}{4} \langle \tchi_\pm^I , \tchi_\pm^J \rangle X^{IJ}$ has thus a $t$ dependence.

This field strength is decomposed on invariant and anti-invariant parts under the involution
$\tau_\ltimes$ as
\be
\tcF_\pm = \tN_\pm + \tM_\pm \rlap{\ .}
\ee
\be
%\p_\pm \tcV \tcV^{-1}
%+ \G[ \p_\pm f \!\circ\! f^{-1} \p_t, \tcV \D] \tcV^{-1} 
\tcG_\pm = \tQ_\pm + \tP_\pm
\ee
where $\tN_\pm$ and $\tQ_\pm$ are respectively $\cK$ and $\mK(\mE_9)$ gauge connections whereas $\G( \tM_\pm , \tP_\pm , \hSigma_\pm \D)$ is covariant
with respect to $\cK \ltimes \mK(\mE_9)$. In fact, this decomposition is considered at a
purely algebraic level: $\tQ_\pm$ and $\tP_\pm$ are formal sums of elements of
$\fk(\fe_9)$ and $\fk(\fe_9)^\perp$, without performing summations and similarly for $\tN_\pm$ and $\tM_\pm$.

We define a $\cK \ltimes K(E_9)$-covariant derivative using $\tN_\pm$ and 
$\tQ_\pm$:
\be
\tD_\pm = \p_\pm + \tQ_\pm + \tN_\pm
\rlap{\ .}
\ee

Using this covariant derivative, we define\footnote{Here, $X$ is a capital
  $\chi$.} $\G( \tPsi_{2\pm} , \tX_\pm , \hPsi_\pm \D)$ as
\be
\G(\begin{array}{c}
\tPsi_{2\pm}^I \\
\tX^I_\pm \\
\hPsi_\pm^I
\end{array}\D)
=
\tD_\pm 
\G(\begin{array}{c}
\tpsi_{2\mp}^I \\
\tchi_\mp^I \\
\hpsi^I_\mp
\end{array}\D)
\ee
where the covariant derivative acts as described in section \ref{rep-ferm}:
\be
\tD_\pm
\G(\begin{array}{c}
\tpsi_{2\mp}^I \\
\tchi_\mp^I \\
\hpsi^I_\mp
\end{array}\D)
=
\G(\begin{array}{l}
\p_\pm \tpsi_{2\mp}^I + \G[ \tN_\pm , \tpsi_{2\mp}^I \D] + \tQ_\pm^{IJ}\!\mid_{t=\pm 1} \tpsi_{2\mp}^J + \tzeta_\pm^{I}
\\
\p_\pm \tchi_\mp^I + \G[ \tN_\pm , \tchi_\mp^I \D] + \G[ \tQ_\pm , \tchi_\mp^I \D] + \tQ_\pm^{IJ}\!\mid_{t=\pm 1} \tchi_\mp^J - \G(\tpsi_{2\mp}^{I(t)}t\p_t \tQ_\pm \D)_T + \txi_\pm^I
\\
\p_\pm \hpsi_\mp^I + \frac{1}{2} \omega\!\G(\tQ_\pm , \tchi_\mp^I \D) + \tQ_\pm^{IJ}\!\mid_{t=\pm 1} \hpsi_\mp^J
\end{array}\D)
\ee
where $\tzeta_\pm^I \in \fk$ and $\txi_\pm^I \in \fk(\fe_9)$ may be adjusted to preserve some gauge choice. (The very explicit transformations in the triangular gauge are given in section \ref{rep-ferm}.)

Eventually, we have bosonic fields $\G(f,\tcV,e^{\hsigma}\D)$ with left and right ``field strength'' 
\be
\G(f,\tcV,e^{\hsigma}\D) \ \longrightarrow \ 
\begin{array}{l}
\G(\tcF_-,\tcG_-,\hSigma_-\D)
\\
\G(\tcF_+,\tcG_+,\hSigma_+\D) \rlap{\ .}
\end{array}
\ee
Similarly, fermionic fields $\G( \tpsi_{2\pm} , \tchi_\pm , \hpsi_\pm \D)$ have left and right ``field strength''
\be
\G( \tpsi_{2\pm} , \tchi_\pm , \hpsi_\pm \D) \ \longrightarrow \ 
\begin{array}{l}
\G( \tpsi_{2\mp} , \tchi_\mp , \hpsi_\mp \D)
\\
\G( \tPsi_{2\pm} , \tX_\pm , \hPsi_\pm \D) \rlap{\ .}
\end{array}
\ee
For bosons, there are two derivatives. For fermions, there is one derivative and one ($d=2$) $\gamma$ matrix which changes the chirality.

A selfduality equation is written in the same way as in \cite{Paulot:2004hh} where it was done for the bosonic sector. Bosonic and fermionic ``Field strength'' are decomposed respectively on 
\be
\fk \ltimes \fk(\fe_9) \ \oplus \ (\fk \ltimes \fk(\fe_9))^\perp
\ee
and
\be
\mathbf{16}_\pm \times_T (\fk \ltimes \fk(\fe_9)) \ \oplus \ \mathbf{16}_\pm \times_T (\fk \ltimes \fk(\fe_9))^\perp \rlap{\ .}
\ee
We denote collectively this decomposition by
\be
\begin{array}{c}
\G(\tcF_\pm,\tcG_\pm,\hSigma_\pm\D)
\\
\G( \tpsi_{2\pm} , \tchi_\pm , \hpsi_\pm \D)
\\
\G( \tPsi_{2\pm} , \tX_\pm , \hPsi_\pm \D)
\end{array}
= \ \tcX_\pm + \tcY_\pm
\ee
where $\tcX_\pm$ is the ``gauge'' compact part and $\tcY_\pm$ is the covariant noncompact part.

The duality operator $\cS$ is
\be
\cS : \
\G\{
\begin{array}{rcl}
\alpha_\pm \, L_n & \longrightarrow & \pm  \alpha_\pm \, L_{1-n} \\
\beta_\pm \, t^nT & \longrightarrow & \mp  \beta_\pm \, t^{1-n} \tau(T) \\
\gamma_\pm \, c & \longrightarrow & - \, \gamma_\pm \, c
\end{array}
\D.
\ee
where $T$ is any generator of $\fe_8$. For fermions, the index $I$ corresponding to the $\mathbf{16}_\pm$ factor is left invariant.
Using this operator, we can write a selfduality equation
\be
\cS \tcY = \tcY
\rlap{\ .}
\label{sd}
\ee
We claim the following
\begin{enumerate}
\item This equation is preserved by all gauge transformations: it is covariant under the gauge action.
\item This equation reduces the infinite-dimensional fields to the fields of maximal $d=2$ supergravity with their equations of motion.
\end{enumerate}
These two points are the objects of the next two sections.

\subsection{Solution of the selfduality equation and gauge-covariance}

The solution of the selfduality equation is obtained on the same line as in section \ref{bsm-deriv}. Let us start with the bosonic field $\tcF$. It can expanded in $t$ as
\be
\tcF_\pm(t) = \sum_{n \in \ZZ} t^n \cF^{(n)}_\pm t\p_t \rlap{\ .}
\ee
The noncompact part is the part anti-invariant under the involution $\tau_\fw$ described in section \ref{k-action}:
\be
\frac{1}{2} \sum_{n\in \ZZ} t^n \G( \cF^{(n)}_\pm + \cF^{(-n)}_\pm \D) t\p_t \rlap{\ .} 
\ee
The selfduality equation (\ref{sd}) is solved for this field as
\be
\frac{1}{2} \G( \cF^{(n)}_\pm + \cF^{(-n)}_\pm \D) = (\mp)^n M_\pm
\ee
which reduces the infinite tower of fields to a single field $M_\pm$.
In exactly the same way, the solutions for the fermionic partners 
\be
\tpsi^I_{2\pm} = \sum_{n \in \ZZ} t^n \psi^{I(n)}_{2\pm} t\p_t 
\ee 
and 
\be
\tPsi^I_{2\pm} = \sum_{n \in \ZZ} t^n \Psi^{I(n)}_{2\pm} t\p_t 
\ee 
are
\be
\frac{1}{2} \G( \psi^{I(n)}_{2\pm} + \psi^{I(-n)}_{2\pm} \D) = (\mp)^n \psi^I_{2\pm}
\ee
\be
\frac{1}{2} \G( \Psi^{I(n)}_{2\pm} + \Psi^{I(-n)}_{2\pm} \D) = (\mp)^n \Psi^I_{2\pm}
\rlap{\ .}
\ee

For matter fields, the bosonic field $\tcG_\pm$ is expanded as
\be
\tcG_\pm(t) = \sum_{n \in \ZZ} t^n \cG^{(n)}_\pm
\ee
where $\cG^{(n)}$ are 1-forms with value in the Lie algebra $\fe_8$.
The noncompact part is
\be
\frac{1}{2} \sum_{n\in \ZZ} t^n \G( \cG^{(n)}_\pm - \tau\!\G(\cG^{(-n)}_\pm\D) \D)
\ee
according to the definition of the involution $\ttau$ given in section \ref{definitions}. $\tau$ is the involution preserving the maximal compact subalgebra $\fso(16)$ of $\fe_8$.
The solution to the selfduality equation reads for these fields reduces again the infinite tower of fields to a single field:
\be
\frac{1}{2} \G( \cG^{(n)}_\pm - \tau\!\G(\cG^{(-n)}_\pm\D) \D) = (\pm)^n P_\pm
\rlap{\ .}
\ee
The case $n=0$ in this formula tells that $P_\pm$ must be noncompact:
\be
P_\pm = P^A_\pm Y^A
\rlap{\ .}
\ee 
Similarly the infinite tower of fermionic fields
\be
\tchi^I_\pm(t) = \sum_{n \in \ZZ} t^n \chi^{I(n)}_\pm
\ee
\be
\tX^I_\pm(t) = \sum_{n \in \ZZ} t^n X^{I(n)}_\pm
\ee
has for solution to (\ref{sd})
\be
\frac{1}{2} \G( \chi^{I(n)}_\pm - \tau\!\G(\chi^{I(-n)}_\pm\D) \D) = (\mp)^n \chi^I_\pm
\ee
\be
\frac{1}{2} \G( X^{I(n)}_\pm - \tau\!\G(X^{I(-n)}_\pm\D) \D) = (\mp)^n X^I_\pm
\ee
with
\be
\chi_\pm^I = \chi_\pm^{I,A} Y^A
\ee
and the analogous for $X_\pm^I$.
Because of the truncation which enters the definition of the fermionic representation, and due to the fact that there is no contribution of the compact part of $\tchi_\pm^I$ in the $Y_A$ at $t=\mp 1$, $\chi_\pm^I$ is in fact a Majorana-Weyl fermion:
with
\be
\chi_\pm^I = \Gamma^I_{A\dot{A}} \chi^{\dot{A}} Y^A
\ee
and the same for the covariant derivative $X^I_\pm$.

Finally, for the fields associated to the central charge the selfduality equation (\ref{sd}) reads
\be
\hSigma = 0
\ee
\be
\hpsi^I_\pm = 0
\ee
\be
\hPsi^I_\pm = 0
\rlap{\ .}
\ee

Let us analyse the different gauge symmetries of the model. The begin by the linear fermionic gauge transformations. $\tpsi_{2\pm}^I$ has values in $\fw/\fk'$: it is defined modulo the gauge transformations
\be
\tpsi_{2\pm}^I \ \longrightarrow \ \tpsi_{2\pm}^I + \tzeta_{\pm}^I
\label{cg-tzeta}
\ee
for $\tzeta_{\pm}^I \in \mathbf{16}_\pm \times \fk_\pm$ where $\fk_\pm$ are element of $\fk$ with vanishing derivative at $t=\mp 1$.  As this transformation changes only the compact part of $\tpsi_{2\pm}^I$, it has no effect on the 
noncompact part and therefore on the selfduality condition for this field. However, $\tpsi_{2\pm}^I$ appears also in the definitions of $\tcF$, $\tcG$, $\tPsi^I_{2}$ and $\tX^I$. In fact, with the condition that the derivative of $\tzeta_\pm$ vanishes at $t=\mp 1$, it can be checked that the transformation (\ref{cg-tzeta}) preserves the selfduality condition and leaves all reduced fields $M_\pm$, $P_\pm$, $\psi^I_2$, $\Psi^I_{2\pm}$, $\chi^I_\pm$ and $X^I_\pm$ invariant. (The effect of a $\tzeta_\pm^I$ with nonvanishing derivative at $t=\mp 1$ will be studied in section \ref{section-susy}.)

Similarly, the transformation
\be
\tchi_\pm^I \ \longrightarrow \ \tchi_\pm^I + \txi_\pm^I
\label{cg-txi}
\ee
does not break the selfduality equation and leave all reduced fields invariant, except $P_\pm$ which transforms as
\be
P_\pm \ \longrightarrow \ P_\pm -2 \psi_{2\pm}^I \p_t \txi_\pm^I \!\mid_{t=\mp 1}
\rlap{\ .}
\ee

The gauge action of $\cK$ for an infinitesimal element $\delta k \in \fk$ preserves also the selfduality condition, with the following transformations of reduced fields:
\bea
M_\pm &\longrightarrow& M_\pm - 2\p_t \delta k \!\mid_{t=\mp 1} M_\pm
\\
\psi_{2\pm}^I &\longrightarrow& \psi_{2\pm}^I - 2\p_t \delta k \!\mid_{t=\mp 1} \psi_{2\pm}^I
\\
\Psi_{2\pm}^I &\longrightarrow& \Psi_{2\pm}^I - 2\p_t \delta k \!\mid_{t=\mp 1} \Psi_{2\pm}^I
\\
P_\pm &\longrightarrow& P_\pm - \p_t \delta k \!\mid_{t=\mp 1} P_\pm
\\
\chi_\pm^I &\longrightarrow& \chi_\pm^I - \p_t \delta k \!\mid_{t=\mp 1} \chi_\pm^I
\\
X_\pm^I &\longrightarrow& X_\pm^I - \p_t \delta k \!\mid_{t=\mp 1} X_\pm^I
\rlap{\ .}
\eea

The action of $\mK(\mE_9)$ for an infinitesimal $\delta h \in \fk(\fe_9)$ also preserves the selfduality equation, with the transformations of reduces fields
\bea
M_\pm &\longrightarrow& M_\pm
\\
\psi_{2\pm}^I &\longrightarrow& \psi_{2\pm}^I + \delta h^{IJ}\!\mid_{t=\mp 1} \psi_{2\pm}^J
\\
\Psi_{2\pm}^I &\longrightarrow& \Psi_{2\pm}^I + \delta h^{IJ}\!\mid_{t=\mp 1} \Psi_{2\pm}^J
\\
P_\pm &\longrightarrow& P_\pm + \G[ \delta h\!\mid_{t=\mp 1} , P_\pm\D] - M_\pm (t\p_t \delta h)\!\mid_{t=\mp 1}
\\
\chi_\pm^I &\longrightarrow& \chi_\pm^I + \G[ \delta h\!\mid_{t=\mp 1} , \chi_\pm^I \D] + \delta h^{IJ}\!\mid_{t=\mp 1} \chi_\pm^J - \G(  \psi_{2\pm}^I (t\p_t \delta h)\!\mid_{t=\mp 1}\D)_T 
\label{transfo-chi-I}
\\
X_\pm^I &\longrightarrow& X_\pm^I + \G[ \delta h\!\mid_{t=\mp 1} , X_\pm^I \D] + \delta h^{IJ}\!\mid_{t=\mp 1} X_\pm^J - \G(  \Psi_{2\pm}^I (t\p_t \delta h)\!\mid_{t=\mp 1}\D)_T 
\label{transfo-Chi-I}
\rlap{\ .}
\eea
As $\chi_\pm^I = \Gamma^I_{A\dot{A}} \chi^{\dot{A}} Y^A$, the two terms $\G[ \delta h\!\mid_{t=\mp 1} , \chi_\pm^I \D] + \delta h^{IJ}\!\mid_{t=\mp 1} \chi_\pm^J$ reduces in fact to the transformation for a Majorana-Weyl fermion $\frac{1}{4} \delta h^{IJ}\!\mid_{t=\mp 1} \Gamma^{IJ}_{\dot{A}\dot{B}}\chi_\pm^{\dot{B}}$. In addition, the projection $T$ on the Majorana-Weyl representations can be explicitely written as
\be
\G(  \psi_{2\pm}^I (t\p_t \delta h)\!\mid_{t=\mp 1}\D)_T = \psi_{2\pm}^J \G(\Gamma^I \Gamma^J\D)_{AB} (t\p_t \delta h^B)\!\mid_{t=\mp 1} Y^A \rlap{\ .}
\ee
The transformation (\ref{transfo-chi-I}) and similarly (\ref{transfo-Chi-I}) can thus be rewritten in term of $\chi_\pm^{\dot{A}}$ and $X_\pm^{\dot{A}}$ as
\be
\chi_\pm^{\dot A} \ \longrightarrow \ \chi_\pm^{\dot A} + \frac{1}{4} \delta h^{IJ}\!\mid_{t=\mp 1} \Gamma^{IJ}_{\dot{A}\dot{B}}\chi_\pm^{\dot{B}} \pm \psi_{2\pm}^I \Gamma^I_{A\dot{A}} \p_t \delta h^A\!\mid_{t=\mp 1}
\ee
and
\be
X_\pm^{\dot A} \ \longrightarrow \ X_\pm^{\dot A} + \frac{1}{4} \delta h^{IJ}\!\mid_{t=\mp 1} \Gamma^{IJ}_{\dot{A}\dot{B}}X_\pm^{\dot{B}} \pm \Psi_{2\pm}^I \Gamma^I_{A\dot{A}} \p_t \delta h^A\!\mid_{t=\mp 1}
\ee

We have thus worked out all gauge transformations for the various fields. The selfduality equation is preserved by all transformations and the reduced fields see a very little part of the gauge group $\cK \ltimes \mK(\mE_9)$: only the first derivative $\p_t \delta k \!\mid_{t=\mp 1}$ of the $\cK$ component at $t=\mp 1$ ($\delta k$ vanishes at $t=\mp 1$) and the value $\delta h^{IJ}\!\mid_{t=\mp 1}$ and the first derivative $\p_t \delta h^A\!\mid_{t=\mp 1}$ of the $\mK(\mE_9)$ component at the same point $t=\mp 1$.

\subsection{Linear system in triangular gauge}

The physical content of the theory, with the usual fields, is recovered in triangular gauge. In this gauge, fields are regular in $t$ at $t=0$. This means that $\tcV$ and $\tchi$ are holomorphic at $t=0$ and have a regular Taylor expansion. In particular, the usual physical field $\cV$ is recovered as $\tcV(t=0)$ and similarly for $\tchi$. $f$ leaves the origin invariant: $f(0)=0$ and can be expanded around this point. The dilaton $\rho$ is the first coefficient of this expansion: $\rho = \p_s f(s=0)$. Because it has values in the coset $\fw/\fk_\pm$, $\tpsi_2$ can only be put in the form
\be
\tpsi_2 = \sum_{n \geq -1} \psi_{2(n)} L_{n} = \psi_{2(-1)} t^{n+1} \p_t \rlap{\ .}
\ee

In such a triangular gauge, the solution to the selfduality condition (\ref{sd}) simplify. For example, $\tchi$ has for solution in a general gauge
\be
\chi_{\pm(n)}^I - \tau(\chi_{\pm(-n)}^I) = 2 (\mp)^n \Gamma^I_{A\dot{A}} \chi_\pm^{\dot{A}} Y^A
\rlap{\ .}
\ee
In the triangular gauge, all negative degree components vanish: $\chi_{\pm(-n)}^I=0$ for $n>0$. 
This gives immediately
\be 
\chi_{\pm(n)}^I = 2 (\mp)^n \Gamma^I_{A\dot{A}} \chi_\pm^{\dot{A}} Y^A
\ee 
The expansion reads therefore
\be
\tchi_\pm^I = \Gamma^I_{A\dot{A}} \chi_\pm^{\dot{A}} Y^A \mp 2t \Gamma^I_{A\dot{A}} \chi_\pm^{\dot{A}} Y^A + 2 t^2 \Gamma^I_{A\dot{A}} \chi_\pm^{\dot{A}} Y^A + \ldots 
\ee
and can be summed into
\be
\tchi_\pm^I = \frac{1 \mp t}{1 \pm t} \Gamma^I_{A\dot{A}} \chi_{\dot{A}} Y^A
\rlap{\ .}
\ee

The solution to the selfduality constraint in triangular gauge can be derived for other fields in the same way. The result is the following.
\bea
\tcF_\pm
&=& -4 \G( \frac{1}{t} - t\D) \psi_{2\pm(-1)}^I \psi_{2\pm}^I t\p_t + \frac{1 \mp t}{1 \pm t} \G( \p_\pm \rho \rho^{-1} \pm 16 \psi_{2\pm(-1)}^I \psi_{2\pm}^I\D) t\p_t
\label{ls-1}
\\
\tpsi_{2\pm}^I &=& \G(\frac{1}{t} - t \D) \psi_{2\pm(-1)}^I t\p_t+ \frac{1 \mp t}{1 \pm t} \psi_{2\pm}^I t\p_t
\label{ls-2}
\\
\tPsi_{2\pm}
&=& \G(\frac{1}{t} - t \D) D_\pm \psi_{2\mp(-1)}^I t\p_t+
\frac{1 \mp t}{1 \pm t} D_\pm \psi_{2\mp}^I t\p_t
\label{ls-3}
\eea
\bea
\tcG_\pm
&=&
\G( 
Q_\pm + \frac{1}{4} \chi_\pm^{\dot{A}} \Gamma^{IJ}_{\dot{A}\dot{B}}
 \chi_\pm^{\dot{B}} X^{IJ} -16 \psi_{2\pm(-1)}^I \psi_{2\pm(-1)}^J X^{IJ}
\D)
+ \frac{1 \mp t}{1 \pm t} ( P_\pm \pm 4 \psi_{2\pm(-1)}^I \chi_\pm^I ) \qquad 
\label{ls-4}
\\
\tchi_\pm^I &=& 
\frac{1 \mp t}{1 \pm t} \chi_\pm^{\dot{A}} \, \Gamma^I_{A\dot{A}} Y^A
\label{ls-5}
\\
\tX_\pm^I
&=& 
\frac{1 \mp t}{1 \pm t}  D_\pm \chi_\mp^{\dot{A}} \, \Gamma^I_{A\dot{A}} Y^A
\label{ls-6}
\eea
\bea
\hSigma &=& 0
\label{ls-7}
\\
\hpsi_\pm^I &=& 0
\label{ls-8}
\\
\hPsi_\pm^I &=& 0 \rlap{\ .}
\label{ls-9}
\eea

The compatibility of these different equations gives the equations of motions for all fields. The equations (\ref{ls-7}) and (\ref{ls-8}) are the conformal and superconformal constraints (\ref{eom7}) and (\ref{eom8}). It can be checked that (\ref{ls-9}) is compatible with (\ref{ls-8}).

When the value of $\tpsi_2$ given by (\ref{ls-2}) is used in (\ref{ls-1}), this equation reads
\be
\p_\pm f\!\circ\!f^{-1} \p_t = \frac{1 \mp t}{1 \pm t} \p_\pm \rho \rho^{-1} t\p_t
\ee
which implies equation (\ref{eom3}). Taking into account (\ref{ls-1}) and (\ref{ls-2}), (\ref{ls-3}) reduces to the equation of motion for $\psi^I_{2\pm}$ (\ref{eom4}), up to cubic fermionic terms.

Similarly, with all fields replaced by their values, (\ref{ls-6}) gives the equation of motion for $\chi_\pm^{\dot{A}}$ (\ref{eom2}). Replacing fermionic fields by their explicit solutions (\ref{ls-2}) and (\ref{ls-5}), $\tcG$ can be expressed as
\be
\begin{split}
\tcG_\pm = &\p_\pm \tcV \tcV^{-1} + \p_\pm f\!\circ\!f^{-1} \p_t \tcV \tcV^{-1} 
\\&
\pm 4 \frac{1 \mp t}{1 \pm t} \psi_{2\pm(-1)}^I \Gamma^I_{A\dot{A}} \chi_\pm^{\dot{A}} Y^A 
\pm 4 \frac{t(1 \mp t)}{(1 \pm t)^3} \psi_{2\pm}^I \Gamma^I_{A\dot{A}} \chi_\pm^{\dot{A}} Y^A 
\\&
+ \frac{1}{4} \G( \frac{1 \mp t}{1 \pm t} \D)^2 \chi_\pm^{\dot{A}} \Gamma^{IJ}_{\dot{A}\dot{B}} \chi_\pm^{\dot{B}} X^{IJ}
\\
&- 16 \psi_{2\pm(-1)}^I \psi_{2\pm(-1)}^J X^{IJ}
- 32 \frac{t}{(1 \pm t)^2} \psi_{2\pm(-1)}^I \psi_{2\pm}^J X^{IJ}
- 16 \frac{t^2}{(1 \pm t)^4} \psi_{2\pm}^I \psi_{2\pm}^J X^{IJ}
\rlap{\ .}
\end{split}
\ee
With this expression on the left handside, equation (\ref{ls-4}) takes the form of the linear system introduced in \cite{Nicolai:1988jb}, with $\psi^I_\pm$ replaced here by $4\psi^I_{2\pm(-1)}$. Its compatibility conditions gives in the usual way the equation of motion (\ref{eom1}) for matter bosons $\cV$ but also all equations for fermions, according to \cite{Nicolai:1998gi}. Furthermore, these equations are exact and include all higher order fermionic conditions. In particular, it implies the equation for $4\psi^I_{2\pm(-1)}$, which replaces here $\psi^I_\pm$:
\be
4\psi^I_{2\pm(-1)} = 
 -2 \frac{\p_\pm \sigma \,  \rho \psi^I_{2\pm}}{\p_\pm \rho} 
+ \frac{D_\pm(\rho \psi^I_{2\pm})}{\p_\pm \rho} 
+ \frac{\langle P_\pm , \tchi_\pm^I \rangle}{\rho^{-1} \p_\pm \rho}
+ \ldots
\label{psi-1}
\ee
Here, $\psi^I_\pm$ appears in the definition of $\hsigma$ and $\hpsi$. In particular, equation (\ref{ls-8}) reads in term of usual fields
\be
\psi^I_\pm = 
 -2 \frac{\p_\pm \sigma \,  \rho \psi^I_{2\pm}}{\p_\pm \rho} 
+ \frac{D_\pm(\rho \psi^I_{2\pm})}{\p_\pm \rho} 
+ \frac{\langle P_\pm , \tchi_\pm^I \rangle}{\rho^{-1} \p_\pm \rho}
+ \ldots
\ee
which is exactly the same equation as (\ref{psi-1}). As a consequence, the compatibility of the full set of equations (\ref{ls-1})~--~(\ref{ls-9}) implies
\be
\psi^I_{2\pm(-1)} = \frac{1}{4}\psi^I_\pm \rlap{\ .}
\ee
Finally, equation (\ref{ls-7}) is the conformal constraint (\ref{eom7}).

\subsection{Supersymmetry}
\label{section-susy}

Supersymmetry transformations can be defined for the infinite-dimensional fields. Up to higher order fermionic terms, the supersymmetry generators act on bosons as
\be 
\delta_\pm \G( f , \tcV , e^{\hsigma}\D) \G( f , \tcV , e^{\hsigma} \D)^{-1} = \epsilon_\pm^I \G( \tpsi_{2\pm}^I , \tchi_\pm^I , \hpsi_\pm^I\D)
\ee
which reads explicitely
\be
\begin{array}{rcl}
\delta_\pm f\!\circ\! f^{-1} \p_t &=& \pm 2 \epsilon_\pm^I \tpsi_{2\pm}^I
\\
\delta_\pm \tcV \tcV^{-1} + \delta_\pm f\!\circ\! f^{-1} \p_t \tcV \tcV^{-1} &=& - 2 \epsilon_\pm^I \tchi_\pm^I 
\\
\delta_\pm \hsigma + \Omega'\!\G(\tcV, \delta_\pm \tcV \tcV^{-1} + \delta_\pm f\!\circ\! f^{-1} \p_t \tcV \tcV^{-1} \D) &=& - 2 \epsilon_\pm^I \hpsi_\pm^I
\end{array}
\ee
up to various gauge transformations.
For fermions, the transformations are, at the first order in fermions,
\be
\begin{array}{rcl}
\delta_\pm \tpsi_{2\pm}^I &=& \frac{1}{4} \G( \frac{1}{t} -t \D) \G( \tD_\pm \epsilon_\pm^I + \p_\pm \hsigma \epsilon_\pm^I \D) t\p_t
\mp \epsilon_\pm^I \tcF_\pm 
\\
\delta_\pm \tchi_\pm^I &=& \G(\epsilon_\pm^I \tcG_\pm\D)_T 
\\
\delta_\pm \hpsi_\pm^I &=& \epsilon_\pm^I \hSigma_\pm
\end{array}
\ee
up to additive gauge transformations for $\tpsi_{2\pm}^I$ and $\tchi_\pm^I$. These supersymmetry transformations preserve the selfduality condition and therefore the equations of motion for the physical fields.

\subsection*{Aknowledgments}
This work is partially supported by IISN - Belgium (convention
4.4505.86), by the ``Interuniversity Attraction Poles Programme --
Belgian Science Policy'' and by the European Commission FP6
programme MRTN-CT-2004-005104, in which we are associated to
V.U.Brussel.

\end{document}